%% file: RSPup_pFactor_v3r0.tex
\documentclass[traditabstract]{aa}
\usepackage{graphics}
\usepackage{graphicx}
\usepackage{multicol}
\usepackage{longtable}
\usepackage{supertabular}
\usepackage{txfonts}

\usepackage{natbib,twoopt}
\usepackage{amsmath} 
\usepackage[breaklinks=true]{hyperref} 
\bibpunct{(}{)}{;}{a}{}{,} 
\makeatletter
\newcommandtwoopt{\citeads}[3][][]{\href{http://adsabs.harvard.edu/abs/#3}%
{\def\hyper@linkstart##1##2{}%
\let\hyper@linkend\@empty\citealp[#1][#2]{#3}}}
\newcommandtwoopt{\citepads}[3][][]{\href{http://adsabs.harvard.edu/abs/#3}%
{\def\hyper@linkstart##1##2{}%
\let\hyper@linkend\@empty\citep[#1][#2]{#3}}}
\newcommandtwoopt{\citetads}[3][][]{\href{http://adsabs.harvard.edu/abs/#3}%
{\def\hyper@linkstart##1##2{}%
\let\hyper@linkend\@empty\citet[#1][#2]{#3}}}
\newcommandtwoopt{\citeyearads}[3][][]%
{\href{http://adsabs.harvard.edu/abs/#3}
{\def\hyper@linkstart##1##2{}%
\let\hyper@linkend\@empty\citeyear[#1][#2]{#3}}}
\makeatother

\usepackage{color}
\definecolor{mygreen}{RGB}{0,128,0}

\hypersetup{colorlinks=true,linkcolor=blue,citecolor=blue,urlcolor=blue}

\def\pfactorRSPup{1.250}
\def\pfactorRSPuperrstat{0.034} 
\def\pfactorRSPuperrsyst{0.054} 
\def\pfactorRSPuperrtot{0.064}
\def\relerrpfactor{5.1} 
\def\tetzero{0.8490}
\def\tetzeroerrstat{0.0034}
\def\tetzeroerrsyst{0.0120}
\def\tetmean{0.9305}
\def\tetmeanerrstat{0.0034}
\def\tetmeanerrsyst{0.0120}
\def\EBV{0.4961}
\def\EBVerr{0.0060}
\def\excessK{0.027}
\def\excessKerr{0.011}
\def\excessH{0.016}
\def\excessHerr{0.011}
\def\vgamma{$25.423 \pm 0.200$}

\def\MJDzero{45\,838.0313}
\def\JDzeroerr{0.098}
\def\Pzero{41.438138}  
\def\Pzeroerr{0.00070}
\def\PERIODone{-3.42244 \times 10^{-7}}
\def\PERIODtwo{+0.23085 \times 10^{-8}}
\def\PERIODthree{+0.13219 \times 10^{-12}}
\def\PERIODfour{-0.74919 \times 10^{-16}}
\def\PERIODfive{+0.42849 \times 10^{-20}}
\def\Pchange{+114.8} 
\def\meanRadius{$191$}
\def\ampRadius{$164 / 208$} 
\def\meanTeff{$5060$} 
\def\ampTeff{$4640 / 5850$} 
\def\meanLuminosity{$21\,700$}
\def\ampLuminosity{$14\,200 / 29\,500$}
\def\meanBolometricmag{$-6.072$}
\def\ampBolometricmag{$-6.434$/$-5.640$}
\def\pfactorLCar{1.27} 
\def\pfactorLCarerr{0.12}
\def\distRSPup{1910} 
\def\distRSPuperr{80}
\def\meanp{1.293}
\def\meanpwithFFAql{1.285}
\def\meanperr{0.039}
\def\meanprelerr{3.0} 
\def\pchired{0.9}
\begin{document}

\title{Observational calibration of the projection factor of Cepheids}
\subtitle{III. The long-period Galactic Cepheid RS\,Puppis\thanks{Based on observations collected at the European Organisation for Astronomical Research in the Southern Hemisphere under ESO programs 093.D-0316(A), 094.D-0773(B), 096.D-0341(A) and 098.D-0067(A). Based in part on observations with the 1.3 m telescope operated by the SMARTS Consortium at Cerro Tololo Interamerican Observatory.}}
\titlerunning{Projection factor of the long-period Cepheid RS\,Pup}
\authorrunning{P. Kervella et al.}
\author{
Pierre~Kervella\inst{1,2}
\and
Boris~Trahin\inst{1,2}
\and
Howard~E.~Bond\inst{3}
\and
Alexandre~Gallenne\inst{4}
\and
Laszlo~Szabados\inst{5}
\and
Antoine~M\'erand\inst{4}
\and
Joanne~Breitfelder\inst{2,4}
\and
Julien~Dailloux\inst{1,6}
\and
Richard~I.~Anderson\inst{7,8}
\and
Pascal~Fouqu\'e\inst{9,10}
\and
Wolfgang~Gieren\inst{11}
\and
Nicolas~Nardetto\inst{12}
\and
Grzegorz Pietrzy{\'n}ski\inst{13}
}
\institute{
Unidad Mixta Internacional Franco-Chilena de Astronom\'{i}a (CNRS UMI 3386), Departamento de Astronom\'{i}a, Universidad de Chile, Camino El Observatorio 1515, Las Condes, Santiago, Chile, \email{pkervell@das.uchile.cl}.
\and
LESIA (UMR 8109), Observatoire de Paris, PSL Research University, CNRS, UPMC, Univ. Paris-Diderot, 5 Place Jules Janssen, 92195 Meudon, France, \email{pierre.kervella@obspm.fr}.
\and
Department of Astronomy \& Astrophysics, 525 Davey Lab., Pennsylvania State University, University Park, PA 16802 USA.
\and
European Southern Observatory, Alonso de C\'ordova 3107, Casilla 19001, Santiago 19, Chile.
\and
Konkoly Observatory, MTA CSFK, Konkoly Thege M. \'ut 15-17, H-1121, Hungary.
\and
Institut Sup\'erieur de l'A\'eronautique et de l'Espace, 10 Avenue Edouard Belin, 31400 Toulouse, France
\and
Physics and Astronomy Department, The Johns Hopkins University, 3400 N. Charles St, Baltimore, MD 21218, USA
\and
Observatoire de Gen\`eve, Universit\'e de Gen\`eve, 51 Ch. des Maillettes, 1290 Sauverny, Switzerland.
\and
IRAP, UMR 5277, CNRS, Universit\'e de Toulouse, 14 av. E. Belin, F-31400 Toulouse, France.
\and
CFHT Corporation, 65-1238 Mamalahoa Hwy, Kamuela, Hawaii 96743, USA
\and
Universidad de Concepci{\'o}n, Departamento de Astronom\'{\i}a, Casilla 160-C, Concepci{\'o}n, Chile.
\and
Laboratoire Lagrange, UMR7293, Universit\'e de Nice Sophia-Antipolis, CNRS, Observatoire de la C{\^o}te d'Azur, Nice, France
\and
Nicolaus Copernicus Astronomical Center, Polish Academy of Sciences, ul. Bartycka 18, PL-00-716 Warszawa, Poland.
}
\date{Received ; Accepted}
\abstract
   {The projection factor ($p$-factor) is an essential component of the classical Baade-Wesselink (BW) technique, which is commonly used to determine the distances to pulsating stars.
   It is a multiplicative parameter used to convert radial velocities into pulsational velocities.
   As the BW distances are linearly proportional to the $p$-factor, its accurate calibration for Cepheids is of critical importance for the reliability of their distance scale.
   We focus on the observational determination of the $p$-factor of the long-period Cepheid RS\,Pup ($P=41.5$\,days).
   This star is particularly important as this is one of the brightest Cepheids in the Galaxy and an analog of the Cepheids used to determine extragalactic distances.
   An accurate distance of $1910 \pm 80$\,pc ($\pm 4.2\%$) has recently been determined for RS\,Pup using the light echoes propagating in its circumstellar nebula.
   We combine this distance with new VLTI/PIONIER interferometric angular diameters, photometry, and radial velocities to derive the $p$-factor of RS\,Pup using the code Spectro-Photo-Interferometry of Pulsating Stars (SPIPS). 
   We obtain $p = \pfactorRSPup \pm \pfactorRSPuperrtot\ (\pm \relerrpfactor \%)$, defined for cross-correlation radial velocities.
Together with measurements from the literature, the $p$-factor of RS\,Pup confirms the good agreement of a constant $\overline{p}=\meanp \pm \meanperr\ (\pm \meanprelerr\%)$ model with the observations.
We conclude that the p-factor of Cepheids is constant or mildly variable over a broad range of periods (3.7 to 41.5\,days).   
     }

\maketitle


\section{Introduction}

The oscillation period of Cepheids is longer for more massive, less dense, and more luminous stars. This cyclic change in radius, and its associated effective temperature modulation, is the physical basis of the empirical Leavitt law (the Period-Luminosity relation, \citeads{1908AnHar..60...87L}; \citeads{1912HarCi.173....1L}).
The calibration of the zero-point of the Leavitt law requires the independent measurement of the distances of a sample of Cepheids.
This is complicated by the rarity of these massive stars, and particularly the long-period oscillators, which results in large distances beyond the capabilities of trigonometric parallax measurements.
The parallax-of-pulsation method, also known as the Baade-Wesselink (BW) technique, is a powerful technique to measure the distances to individual Galactic and LMC Cepheids.
The variation of the angular diameter of the star (from surface brightness-color relations or optical interferometry) is compared to the variation of the linear diameter (from the integration of the radial velocity).
The distance of the Cepheid is then obtained by simultaneously
fitting the linear and angular amplitudes (see, e.g.,~\citeads{2011A&A...534A..94S}). 
The main weakness of the BW technique is that it uses a numerical factor to convert disk-integrated radial velocities into photospheric velocities, the projection factor, or $p$-factor (\citeads{2007A&A...471..661N}, \citeads{2009AIPC.1170....3B}, \citeads{2014IAUS..301..145N}).
This factor, whose expected value is typically around 1.3, simultaneously
characterizes the spherical geometry of the pulsating star, the limb darkening, and the difference in velocity between the photosphere and the line-forming regions.
Owing to this intrinsic complexity, the $p$-factor is currently uncertain to 5-10\%, and accounts for almost all the systematic uncertainties of the nearby Cepheid BW distances.
This is the main reason why Galactic Cepheids were excluded from the measurement of $H_0$ by \citetads{2011ApJ...730..119R}.

Recent observational efforts have produced accurate measurements of the $p$-factor of Cepheids \citepads{2005A&A...438L...9M,2013MNRAS.436..953P, 2015A&A...576A..64B, 2015ApJ...815...28G, 2016A&A...587A.117B}, with the objective to reduce this source of systematic uncertainty.
However, these $p$-factor calibrations up to now were essentially obtained on low-luminosity, relatively short-period Cepheids ($P\lesssim10$\,days) that are the most common in the Galaxy.
The most important Cepheids for extragalactic distance determinations are the long-period pulsators ($P \gtrsim 10$\,days), however.
A calibration of the $p$-factor of the intrinsically brightest Cepheids is therefore highly desirable.
Theoretical studies (e.g.,~\citeads{Neilson:2012qy}) indicate that the $p$-factor may vary with the period, but the dependence differs between authors \citepads{Nardetto:2014kx, 2011A&A...534A..94S,2016A&A...587A.117B}.

We focus the present study on the long-period Cepheid \object{RS Pup} (\object{HD 68860}, \object{HIP 40233}, \object{SAO 198944}).
Its period of $P=41.5$\,days makes it one of the brightest Cepheids of our Galaxy and the second nearest long-period pulsator after $\ell$\,Carinae (\object{HD 84810}, $P=35.55$\,days).
\citetads{Kervella:2014lr} reported an accurate measurement of the distance of RS\,Pup, $d = 1910 \pm 80$\,pc, corresponding to a parallax $\pi = 0.524 \pm 0.022$\,mas.
This distance was obtained from a combination of photometry and polarimetry of the light echoes that propagate in its circumstellar dust nebula.
It is in agreement with the \emph{Gaia}-TGAS parallax of $\pi = 0.63 \pm 0.26$\,mas \citepads{2016A&A...595A...2G}, whose systematic uncertainty is estimated to $\pm 0.3$\,mas by \citetads{2016A&A...595A...4L}.
In the present work, we employ the light echo distance of RS\,Pup, in conjunction with new interferometric angular diameter measurements, photometry, and archival data (Sect.~\ref{observations}) to apply the Spectro-Photo-Interferometry of Pulsating Stars (SPIPS) modeling  (Sect.~\ref{rspupspips}).
Through this inverse version of the parallax-of-pulsation technique, we derive its $p$-factor and compare it to the values obtained for $\ell$\,Car and other Cepheids (Sect.~\ref{discussion}).

\section{Observations}\label{observations}

\subsection{Interferometry}

\begin{table*}
 \caption{Characteristics of the calibrators used for the PIONIER observations of RS\,Pup.
 They were selected from the catalogs of \citetads{2010yCat.2300....0L, 2010SPIE.7734E..4EL} (1 to 3) and \citetads{2005A&A...433.1155M} (4).
 For calibrators 1 to 3, we employed the surface brightness color relations calibrated by \citetads{2004A&A...426..297K} to estimate their angular diameters.}
 \label{table:1}
 \centering
 \renewcommand{\arraystretch}{1.2}
 \begin{tabular}{clcccccccc}
  \hline
  \hline
  Number & Name & Sp.\,Type & $m_{B}$ & $m_{V}$ & $m_{H}$ & $m_{K}$ & $\theta_\mathrm{LD}$ & $\theta_\mathrm{UD\ H}$ \\ 
  & & & & & & & (mas) & (mas) \\
  \hline  \noalign{\smallskip}
  1 & \object{HD 67977} & G8III & 7.10 &  6.21  & 4.29 & 4.06 & $0.738 \pm 0.020$ & $0.713 \pm 0.020$ \\
  2 & \object{HD 69002} & K2III & 7.54 &  6.37  & 3.99 & 3.84 & $0.867 \pm 0.020$  & $0.838 \pm 0.020$  \\  
  3 & \object{HD 68978} & G2V & 7.33 &  6.71 & 5.37 & 5.27 & $0.374 \pm 0.008$ & $0.361 \pm 0.008$ \\
  4 & \object{HD 73947} & K2III & 8.48 &  7.09 & 3.95 & 3.88 & $0.863 \pm 0.012$ & $0.834 \pm 0.012$ \\
  \hline
 \end{tabular}
\end{table*}

\begin{figure}[]
\centering
\includegraphics[width=7.5cm]{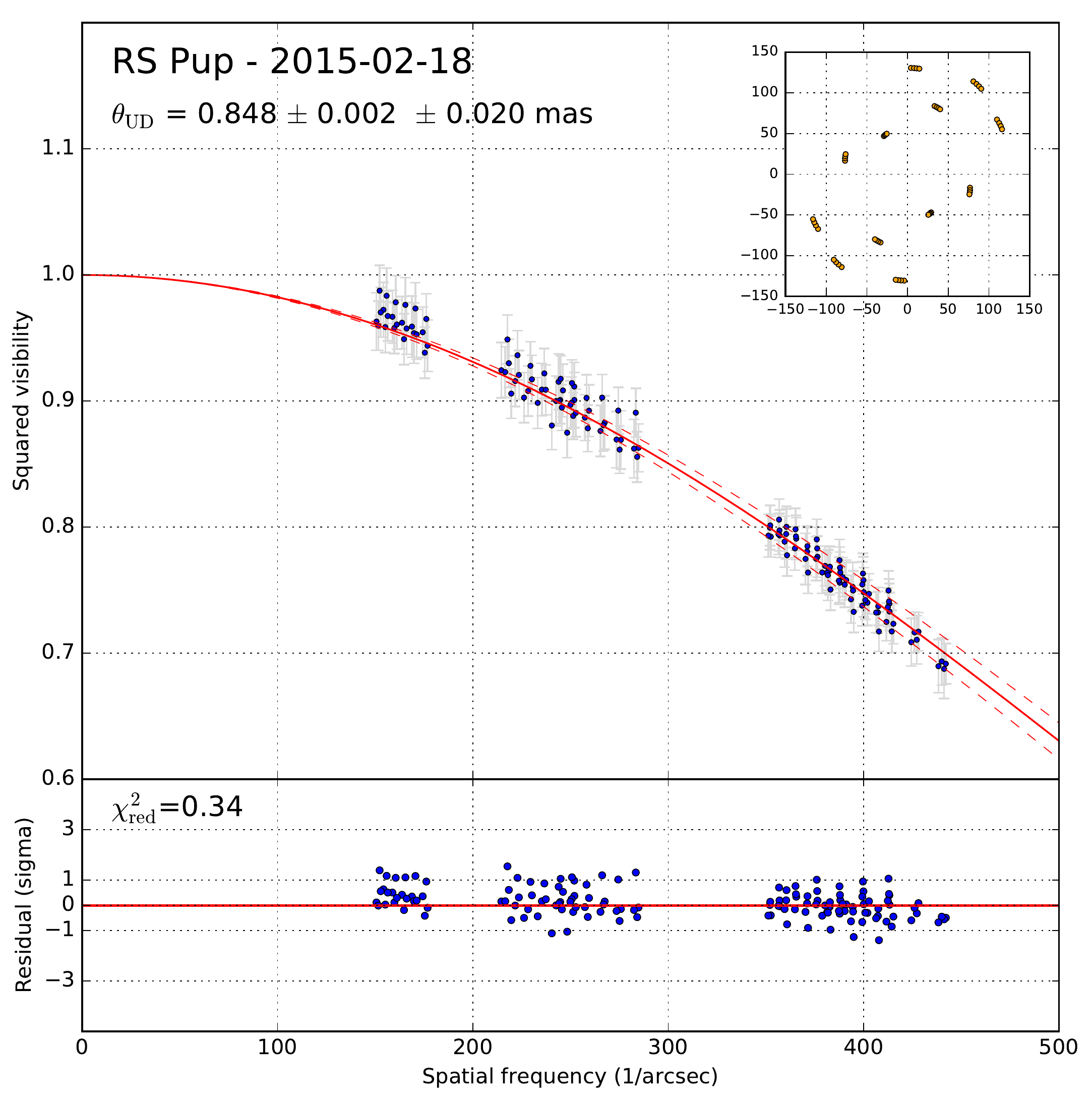}
\includegraphics[width=7.5cm]{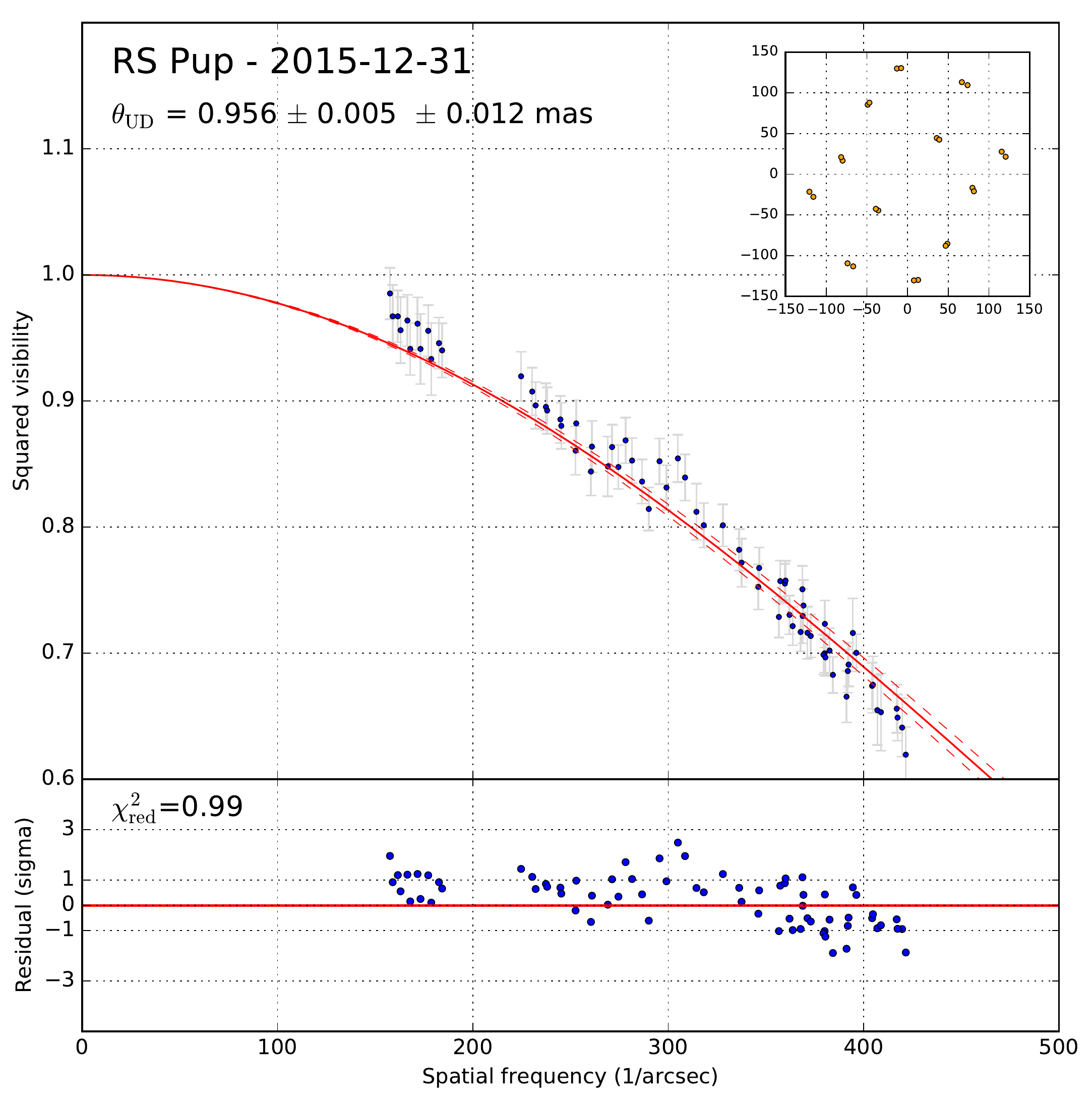}
\caption{Examples of PIONIER squared visibilities collected on RS\,Pup on the night of 18 February 2015, close to the minimum angular diameter phase (top panel), and on 31 December 2015, close to maximum angular diameter (bottom panel).
The solid line is the best-fit uniform disk visibility model, and the dashed lines represent the limits of the $\pm 1 \sigma$ uncertainty domain on the angular diameter.
The $(u,v)$ plane coverage is shown in the upper right subpanels, with axes labeled in meters.\label{visib-PIONIER}}
\end{figure}

We observed RS\,Pup between 2014 and 2016 using the Very Large Telescope Interferometer \citepads{2014SPIE.9146E..0JM} equipped with the PIONIER beam combiner \citepads{2010SPIE.7734E..99B, Le-Bouquin:2011fj}.
This instrument is operating in the infrared $H$ band ($\lambda = 1.6\,\mu$m) using a spectral resolution of $R = 40$.
The four relocatable 1.8\,meter Auxiliary Telescopes (ATs) were positioned at stations A1-G1-J3-K0 or A0-G1-J2-J3\footnote{https://www.eso.org/paranal/telescopes/vlti/configuration/}.
These quadruplets offer the longest available baselines (up to 140\,m), which are necessary to resolve the apparent disk of RS\,Pup ($\theta \approx 0.9$\,mas) sufficiently well.
The pointings of RS\,Pup were interspersed with observations of calibrator stars to estimate the interferometric transfer function of the instrument (Table~\ref{table:1}). These calibrators were selected close angularly to RS\,Pup in order to minimize any possible bias caused by polarimetric mismatch of the beams.
The raw data have been reduced using the {\tt pndrs} data reduction software of PIONIER \citepads{Le-Bouquin:2011fj}, which produces calibrated squared visibilities and phase closures.
Two examples of the measured RS\,Pup squared visibilities are presented in Fig.~\ref{visib-PIONIER}.
The visibilities were classically converted into uniform disk (UD) angular diameters (see, e.g., \citeads{2003AJ....126.2502M} and \citeads{2003EAS.....6..181Y}) that are listed in Table~\ref{table:2}.

\begin{table}
 \caption{PIONIER observations of RS\,Pup. We list the mean modified Julian date (MJD) of each observing night, the calibrator stars, the uniform disk diameter adjusted on the squared visibility measurements, and its statistical and systematic (calibration) uncertainties.}
 \label{table:2}
 \centering
 \renewcommand{\arraystretch}{1.2}
 \begin{tabular}{cccc}
  \hline
  \hline
  UT Date & MJD  & Cal. & $\theta_\mathrm{UD} \pm \sigma_{\rm{stat.}} \pm \sigma_{\rm{syst.}}$\\
   & & & (mas) \\
  \hline  \noalign{\smallskip}
 2014-04-03 & 56750.0178 & 1,2 & $0.860 \pm 0.011 \pm 0.020$ \\ 
 2014-05-08 & 56785.0009 & 1,2 & $0.801 \pm 0.007 \pm 0.020$ \\ 
 2015-01-15 & 57037.3056 & 1,2 & $0.813 \pm 0.005 \pm 0.020$ \\ 
 2015-01-16 & 57038.3660 & 1,2 & $0.830 \pm 0.009 \pm 0.020$ \\ 
 2015-01-17 & 57039.3643 & 1,2 & $0.882 \pm 0.015 \pm 0.020$ \\ 
 2015-02-14 & 57067.1757 & 1,2 & $0.916 \pm 0.005 \pm 0.020$ \\ 
 2015-02-18 & 57071.0997 & 1,2 & $0.848 \pm 0.002 \pm 0.020$ \\ 
 2015-02-21 & 57074.1467 & 1,2 & $0.827 \pm 0.010 \pm 0.020$ \\ 
 2015-12-27 & 57383.1893 & 3,4 & $0.983 \pm 0.007 \pm 0.012$ \\ 
 2015-12-31 & 57387.1858 & 3,4 & $0.956 \pm 0.005 \pm 0.012$ \\ 
 2016-02-21 & 57439.1475 & 3,4 & $0.933 \pm 0.011 \pm 0.012$ \\ 
      \hline
 \end{tabular}
\end{table}

\subsection{Photometry}

\begin{figure}[]
\centering
\includegraphics[width=\hsize]{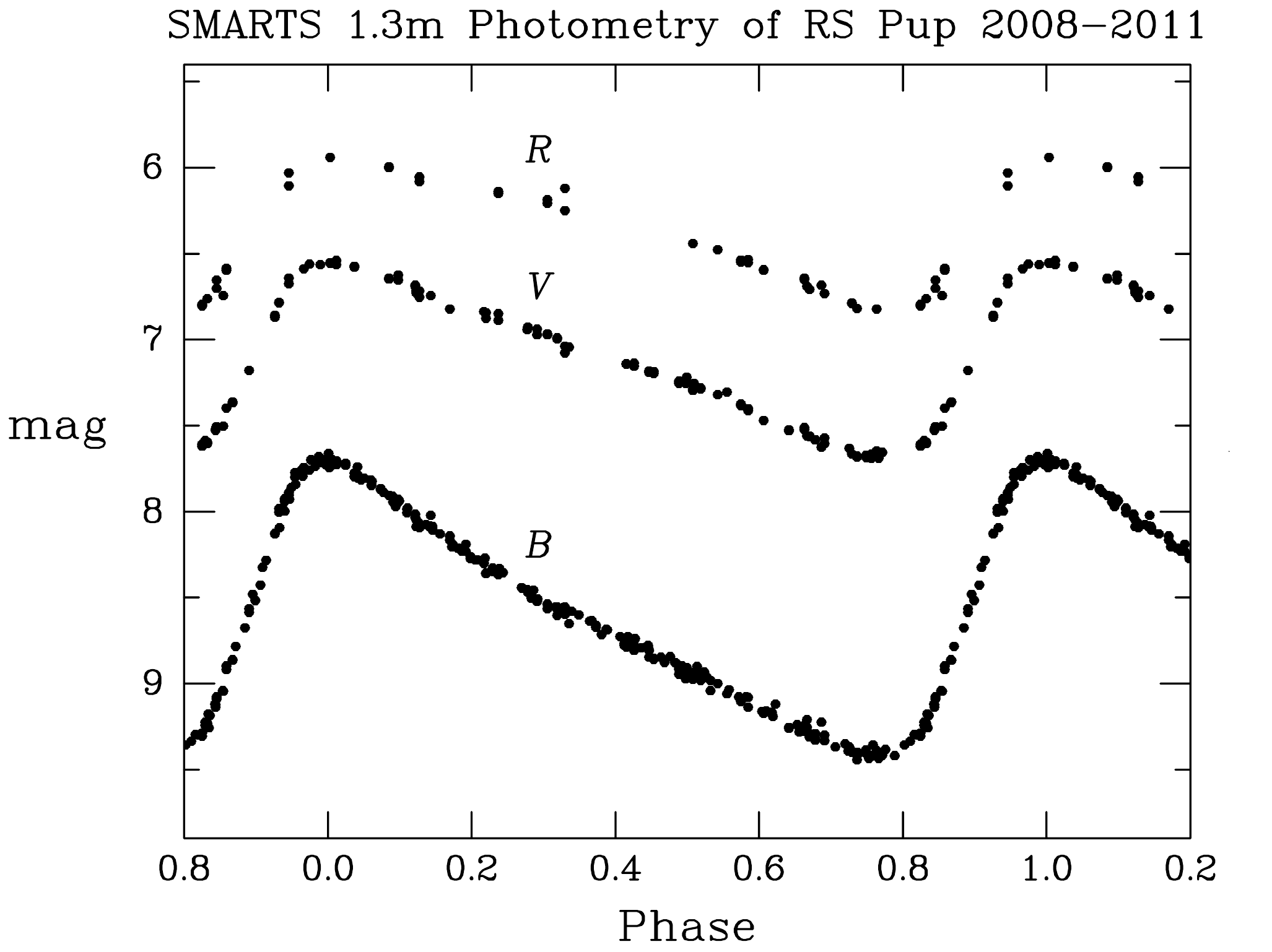}
\caption{SMARTS light curves of RS\,Pup in Johnson $B$, $V$, and Kron-Cousins $R$.\label{SMARTSphotometry}}
\end{figure}

As the measurements available in the literature are of uneven quality for RS\,Pup, we obtained new photometry in the Johnson-Kron-Cousins {\it BVR\/} system using the ANDICAM CCD camera on the SMARTS\footnote{SMARTS is the Small \& Moderate Aperture Research Telescope System; \url{http://www.astro.yale.edu/smarts}} 1.3~m telescope at Cerro Tololo Interamerican Observatory (CTIO)\null.
A total of 277 queue-scheduled observations were made by service observers between 2008 February 28 and 2011 January~25.
The exposure times were usually one second in each filter, but nevertheless, many of the $V$ images and most of the $R$ images were saturated, especially around maximum light, and had to be discarded.
After standard flat-field corrections of the frames, we determined differential magnitudes relative to a nearby comparison star.  In order to convert the relative magnitudes into calibrated values, the {\it BVR\/} magnitudes of the comparison star were determined through observations of \citetads{1992AJ....104..340L} standard-star fields obtained on seven photometric nights.
The resulting {\it BVR\/} light curves are presented in Fig.~\ref{SMARTSphotometry}, phased with a period $P = 41.5113$\,days and $T_0[\mathrm{JD}_\odot] = 2\,455\,501.254$.
As discussed further in this section, this period is suitable over the range of the SMARTS observing epochs (2008-2011).
The list of measured magnitudes is given in Table~\ref{tab-smarts}.
The associated uncertainty is estimated to $\pm 0.03$\,mag per measurement \citepads{2011AJ....141...21W}.

We supplemented the new SMARTS photometric measurements with archival visible light photometry from \citetads{1984ApJS...55..389M}, \citetads{2008yCat.2285....0B}, and \citetads{1976AAS...24..413P}.
In order to improve the coverage of the recent epochs, we also added a set of accurate photoelectric measurements retrieved from the AAVSO database\footnote{\url{https://www.aavso.org}}.
These recent measurements in the Johnson $V$ band are listed in Table~\ref{tab-aavso} and plotted in Fig.~\ref{AAVSOphotometry}.
They cover the JD range 2\,456\,400 (April 2013) to 2\,457\,550 (June 2016), which matches our PIONIER interferometric observations
well.
Finally, we also included in our dataset the near-infrared $JHK$ band photometry from \citetads{Laney:1992fj} and \citetads{1984ApJS...54..547W}.

\subsection{Radial velocities \label{radvel}}

We included in our dataset the radial velocity measurements from \citetads{2014A&A...566L..10A} that provide an excellent coverage of several pulsation cycles of RS\,Pup with a high accuracy.
We complemented these data with the measurements obtained by \citetads{2004A&A...415..531S}.
As discussed by \citetads{2014A&A...566L..10A}, the radial velocity curve of RS\,Pup is not perfectly reproduced cycle-to-cycle.
This is potentially a difficulty for the application of the BW technique, which relies on observational datasets that are generally obtained at different epochs and therefore different pulsation cycles.
This induces an uncertainty on the amplitude of the linear radius variation, and therefore on the derived parameters (distance or $p$-factor).
Following the approach by \citetads{2016MNRAS.455.4231A}, we estimate in Sect.~\ref{spips-pfactor} the uncertainty induced on the $p$-factor by separately fitting the different cycles monitored by \citetads{2014A&A...566L..10A}.

\subsection{Phasing of the datasets}

We took particular care to properly phase the different datasets, a task that is complicated by the rapidly changing period of RS\,Pup.
This is an important step in the fitting process, however, as an incorrect phasing results in biases on the derived model parameters.

\begin{figure*}[]
\centering
\includegraphics[width=8.7cm]{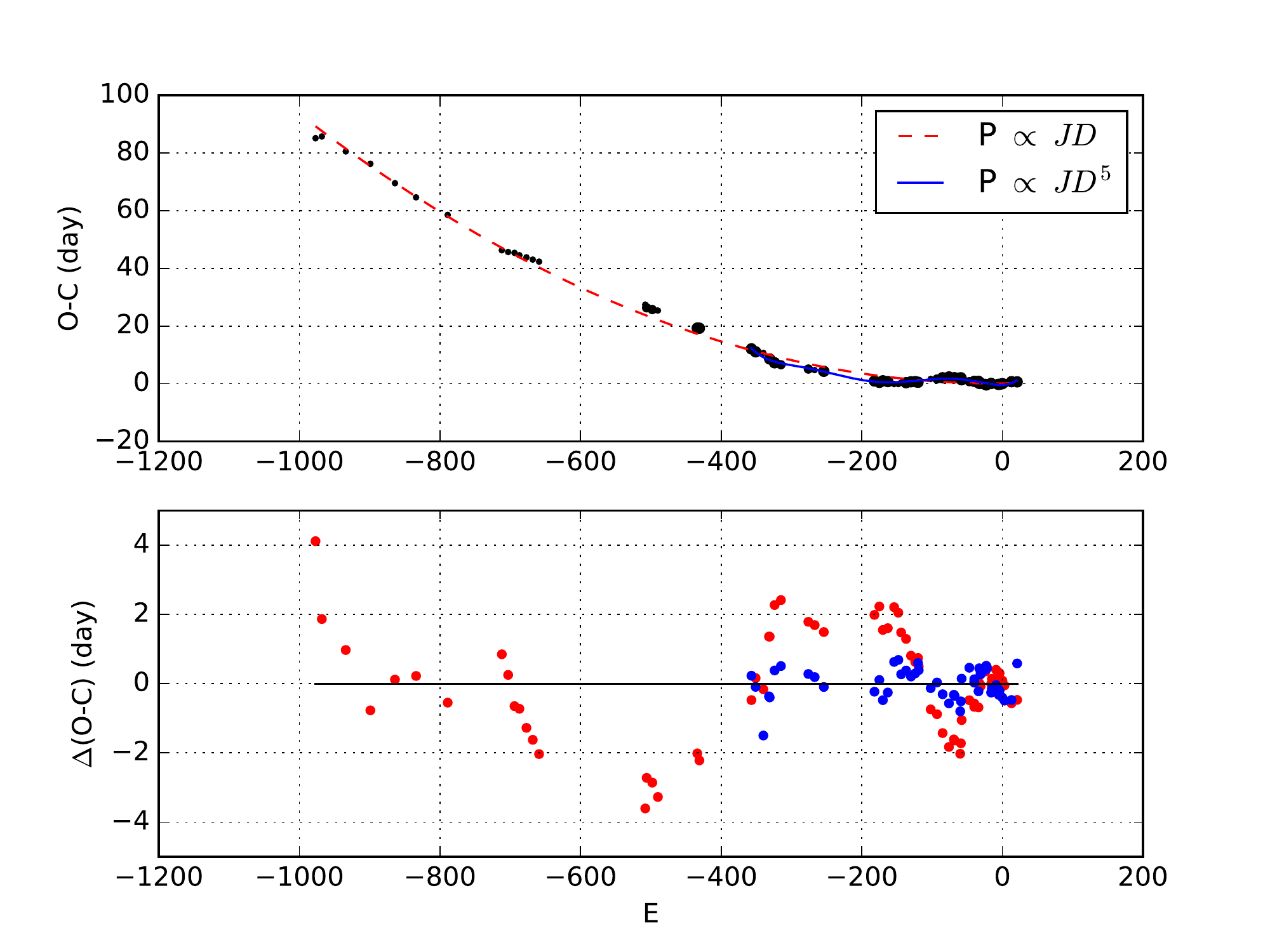}
\includegraphics[width=8.7cm]{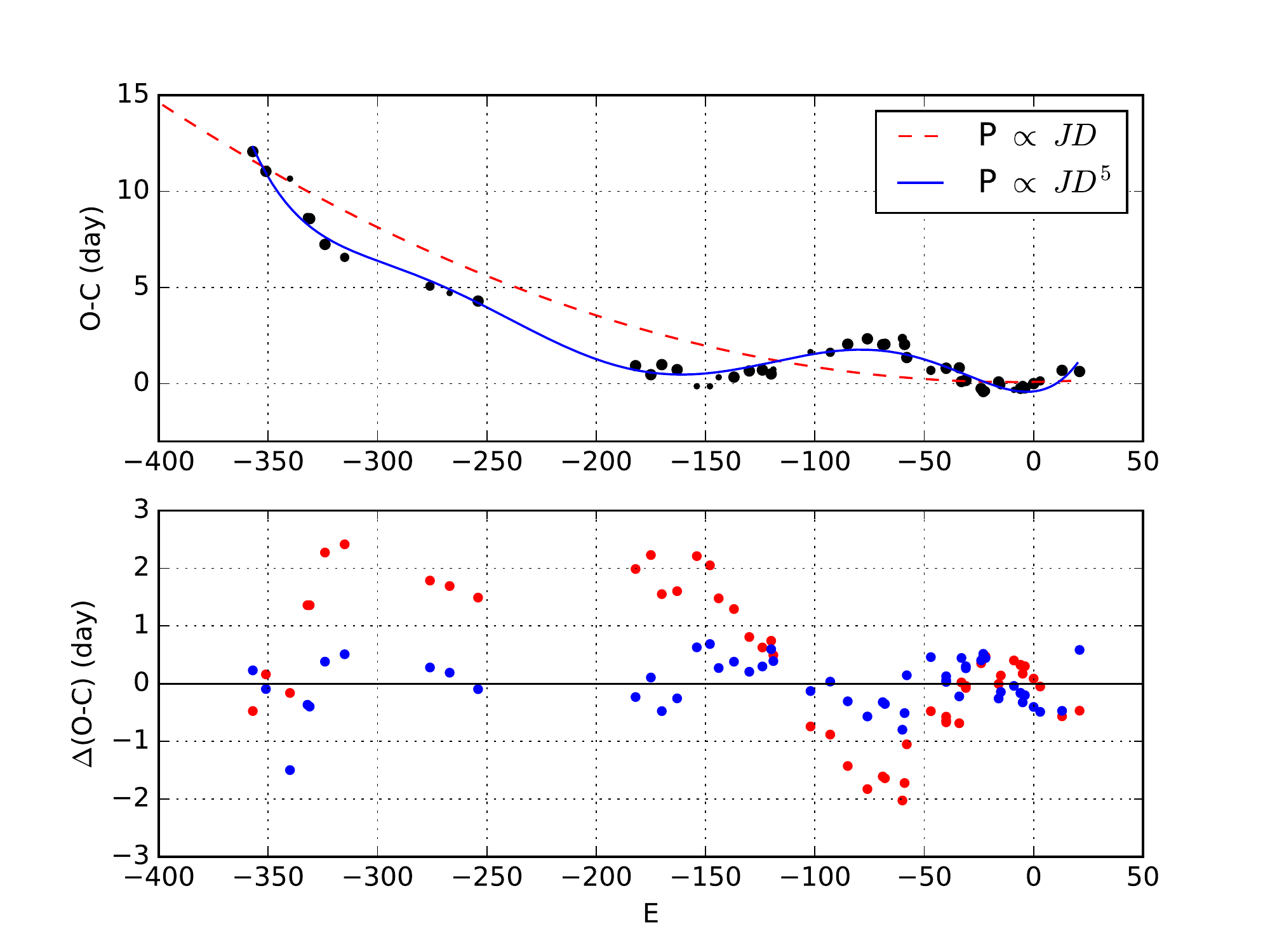}
\caption{{\it Left column:} $O-C$ diagram for RS\,Pup (top panel) and residuals of the model (bottom panel) for a linear period variation (red dashed line and points) and a fifth-degree polynomial function (blue solid line and points). The size of the black points in the upper panel is proportional to their weight in the fit (Table~\ref{tab-oc}). {\it Right column:} enlargement of the $O-C$ diagram covering the last 400 pulsation cycles of RS\,Pup.
\label{OCdiagramRSPup}}
\end{figure*}

As a first-order approach, the pulsation period $P$ and its linear rate of variation have been determined with the classical method of the $O-C$ diagram \citepads{2005ASPC..335....3S}.
The diagram constructed for the moments of the maximum brightness covering more than a century is shown in the left panel of Fig.~\ref{OCdiagramRSPup}.
The relevant data used for constructing the $O-C$ diagram are listed in Table~\ref{tab-oc}.
The general trend of the period variation is an increase, with a superimposed oscillation exhibiting a pseudo-period on the order of three decades.
When calculating the $O-C$ values, the reference epoch $E=0$ was taken as JD$_\odot$~2\,455\,501.254.
This is the normal maximum determined from the SMARTS light curve shown in Fig.~\ref{SMARTSphotometry}.
The variable $E$ designates the number of pulsation cycles that occurred since this reference epoch.
The initial pulsation period was arbitrarily taken as 41.49 days.
The second-order weighted least-squares fit to the $O-C$ residuals is also plotted in Fig.~\ref{OCdiagramRSPup}.
The equation of the fitted parabola is (expressed in Julian date)\begin{equation} \label{linearmodel}
\begin{split}
C =   & \ \  2455501.3428 \pm 0.1756 \\
 & + \Bigl( 41.491734 \pm 0.001404 \Bigr) \times E \\
 &  + \Bigl( 9.515\ 10^{-5} \pm 1.733 \ 10^{-5} \Bigr) \times E^2. \\
\end{split}
\end{equation}
As the $E^2$ coefficient in this equation is positive, the parabola in the $O-C$ diagram tends toward positive values, thus indicating that the period is increasing with time.
Both the $O-C$ graph and the parabolic fit are in good agreement with their counterpart obtained by \citetads{2009AstL...35..406B}, who find a secular period change of $7.824\ 10^{-5} \pm 1.968\ 10^{-5}$ (quadratic term, expressed in fraction of the period per cycle).
The secular period increase that we derive corresponds to a lengthening of +0.1675\,day over a century, or +144.7 s/year.
This value is high, but not without precedent among long-period classical Cepheids \citepads{1980IBVS.1895....1M}.
This rate of secular period change corresponds to the expected value for a third crossing Cepheid with a period like RS\,Pup \citepads{2016A&A...591A...8A}.

The erratic period changes superimposed on the monotonic period variation of RS\,Pup are clearly seen on the residuals of the $O-C$ fit in Fig.~\ref{OCdiagramRSPup}.
In the bottom panels of this figure, the parabola has been subtracted from the $O-C$ values listed in Table~\ref{tab-oc} (as shown in the upper panels).
%
There are three intervals in this diagram where the pulsation period can be approximated with a constant value:
between 1995 and 2002 as $41.518 \pm 0.002$\,days,
between 2003 and 2007 as $41.437 \pm 0.002$\,days, and
between 2008 and 2013 as $41.512 \pm 0.002$\,days.
\citetads{Kervella:2014lr} adopted a period $P=41.5117$\,days for the epoch of the HST/ACS observations (2010) that were used to estimate the distance of RS\,Pup through its light echoes.
It is worth noting that the scatter between the subsequent data points can be intrinsic to the stellar pulsation: this phenomenon is interpreted as a cycle-to-cycle jitter in the pulsation period, as observed in \object{V1154~Cyg}, the only Cepheid in the original Kepler field \citepads{2012MNRAS.425.1312D}.
It was proposed by \citetads{Neilson:2014fk} that the physical mechanism underlying the period jitter of \object{V1154~Cyg} is linked to the presence of convective hot spots on the photosphere of the star.
This explanation may also apply to RS\,Pup, whose relatively low effective temperature could favor the appearance of such convective features.

\begin{figure}[]
\centering
\includegraphics[width=\hsize]{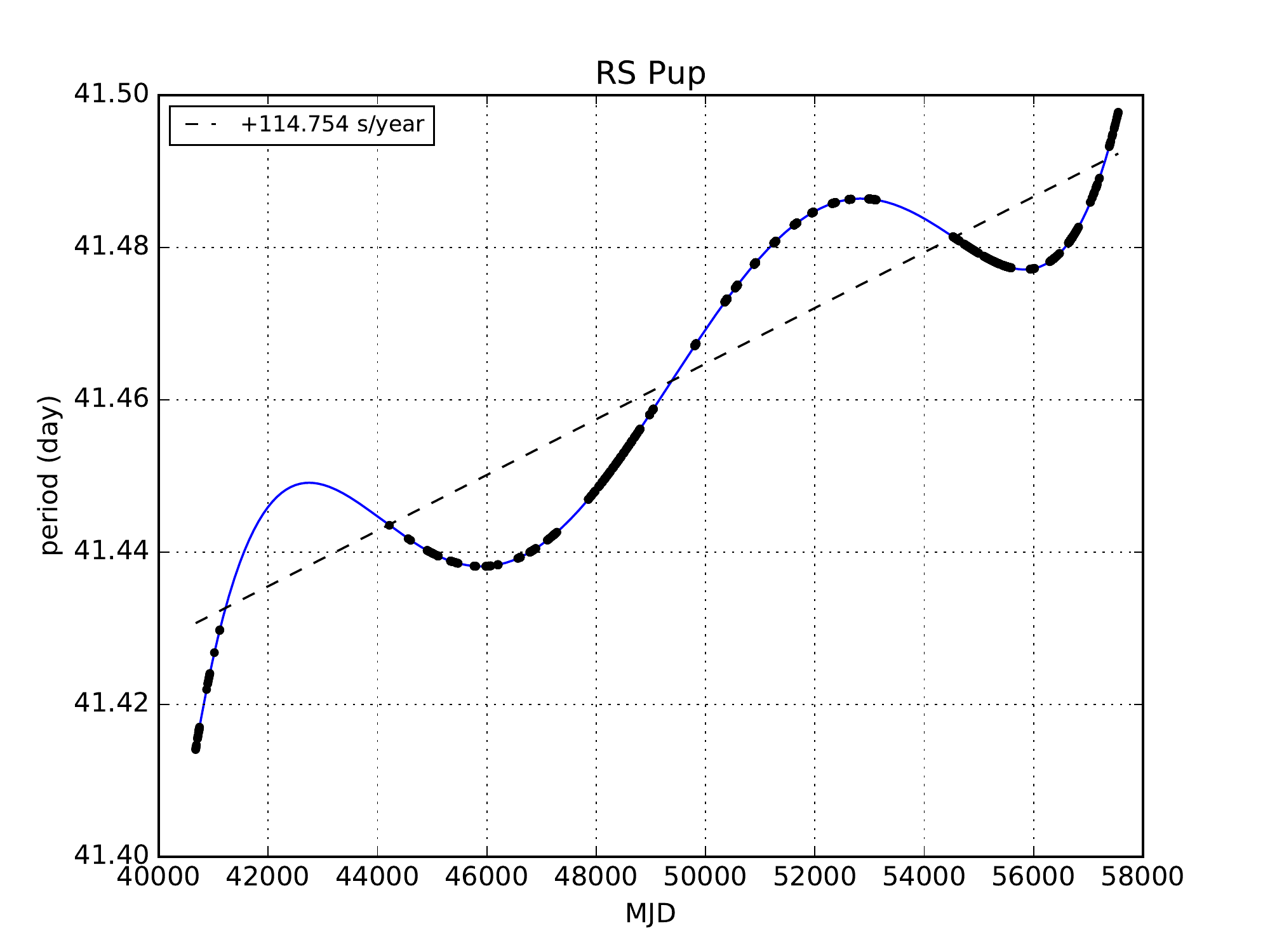}
\caption{Polynomial fit of the changing period of RS\,Pup.
The blue curve is a degree-five polynomial fit of the period values (black points).
The black dashed line represents the linear trend of the period change over the past 50\,years.\label{period_var}}
\end{figure}

The period changes that occurred in the past few decades induced a variability of the maximum light epochs of 3 to 4\,days, that is, up to 0.10 in phase shift.
Such a large phase shift would degrade the quality of the SPIPS combined fit of the observables, in particular the photometry that is spread over four decades.
To take the period changes into account, we adopt a polynomial model of degree five.
This relatively high degree allows us to fit the observed epochs of maximum light much better than the linear model, as shown in the residuals of the $O-C$ diagram (Fig.~\ref{OCdiagramRSPup}, bottom panels).
The period in days as a function of the observing epoch $T$ (expressed in modified Julian date) is given by the polynomial expression
\begin{equation} \label{polymodel}
\begin{split}
P(\mathrm{MJD}) = &\ \  (\Pzero \pm \Pzeroerr)  \PERIODone \, (\Delta t) \\
 &    \PERIODtwo \, (\Delta t)^2 \PERIODthree \,  (\Delta t)^3 \\
 &   \PERIODfour \, (\Delta t )^4 \PERIODfive \, (\Delta t)^5 \\
\end{split}
,\end{equation}
with $\Delta t = \mathrm{MJD} - \mathrm{MJD}_0$ the number of days since the reference epoch $\mathrm{MJD}_0 = \MJDzero$.
The \Pchange\,s/year  linear rate of the period change over the past 50\,years shown in Fig.~\ref{period_var} is close to the value obtained from the fit of the complete dataset with a linearly variable period (+144.7\,s/year).

\section{Analysis of RS\,Pup using SPIPS\label{rspupspips}}

\begin{figure*}[]
\centering
\includegraphics[width=\hsize]{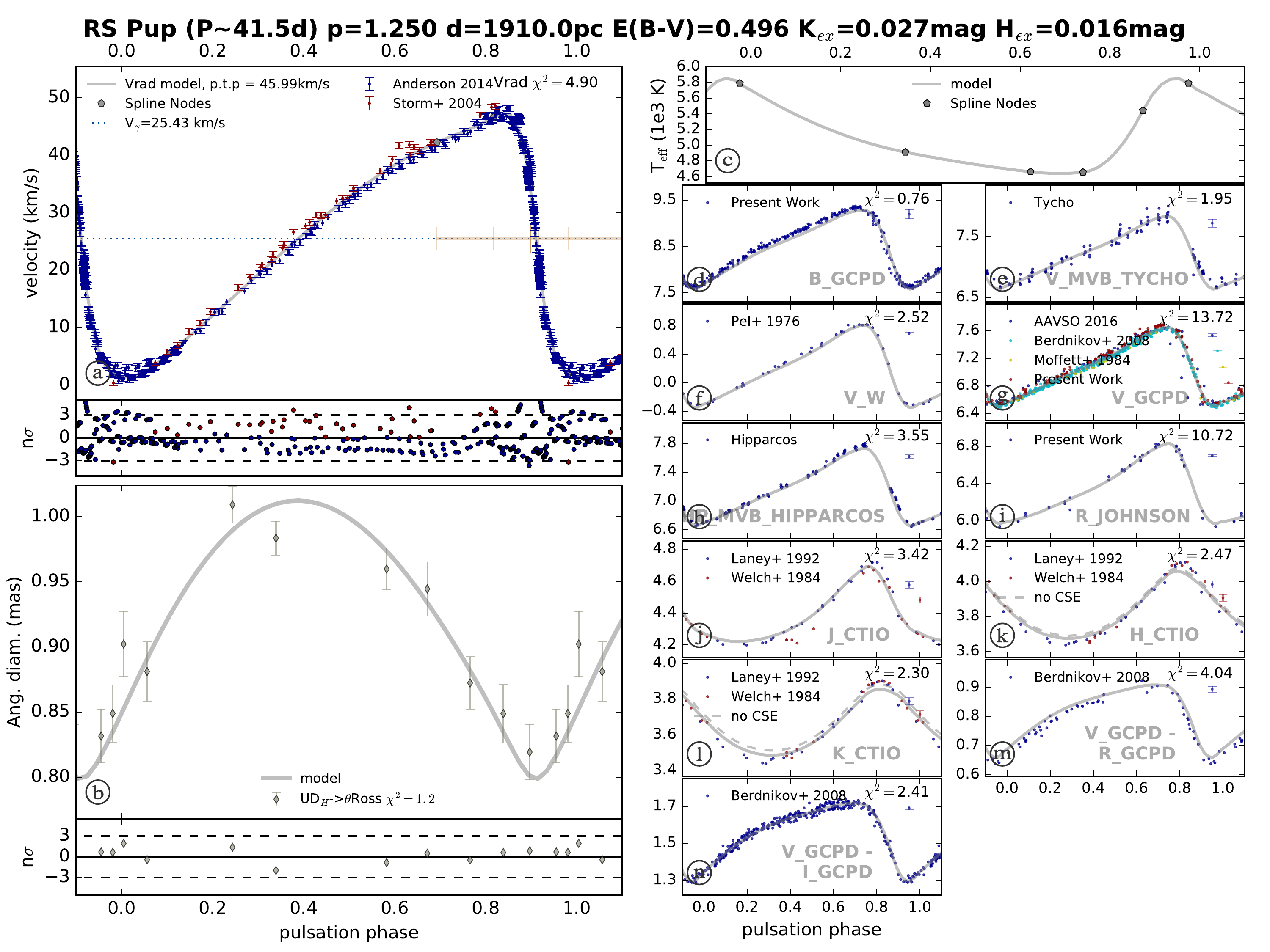} 
\caption{SPIPS combined fit of the observations of RS\,Pup. \label{RSPup-SPIPS}}
\end{figure*}

The SPIPS modeling code \citepads{2015A&A...584A..80M} considers a pulsating star as a sphere with a changing effective temperature and radius, over which is superimposed a combination of atmospheric models from precomputed grids (ATLAS9).
The presence of a circumstellar envelope emitting in the infrared $K$ and $H$ bands is included in the model, as is the interstellar reddening.
The best-fit SPIPS model of RS\,Pup is presented in Fig.~\ref{RSPup-SPIPS} together with the observational data, and the corresponding best-fit parameters are listed in Table~\ref{SPIPS-params}.
The quality of the fit is generally very good for all observing techniques, and the phasing of the different datasets is satisfactory.
The interpolation of the radial velocity curve was achieved using splines with optimized node positions.
We assume  the distance $d=\distRSPup \pm \distRSPuperr$\,pc determined by \citetads{Kervella:2014lr} as a fixed parameter in this fit.

\begin{table}
 \caption{Parameters of the SPIPS model of RS\,Pup. The upper part of the table lists the primary model parameters, and the lower part gives derived physical parameters (mean value and minimum/maximum over the pulsation cycle).}
 \label{SPIPS-params}
 \centering
 \renewcommand{\arraystretch}{1.2}
 \begin{tabular}{ll}
  \hline
  \hline
  Parameter & Value $\pm \sigma_\mathrm{stat} \pm \sigma_\mathrm{syst}$  \\
  \hline  \noalign{\smallskip}
$\theta_0$ (mas)$^a$ & $\tetzero \pm \tetzeroerrstat \pm \tetzeroerrsyst$ \\
$\langle \theta \rangle$ (mas)$^a$ & $\tetmean \pm \tetmeanerrstat \pm \tetmeanerrsyst$ \\
$v_\gamma$ (km\,s$^{-1}$) & \vgamma \\
$E(B-V)$ & $\EBV \pm \EBVerr$ \\
$K$ excess & $\excessK \pm \excessKerr$ \\
$H$ excess & $\excessH \pm \excessHerr$\\
$p$-factor & $\pfactorRSPup \pm \pfactorRSPuperrstat \pm \pfactorRSPuperrsyst$ \\
MJD$_0$ & $\MJDzero \pm \JDzeroerr$ \\
Period (days)$^b$ & $\Pzero \pm \Pzeroerr$ \\
Period change (s/year)$^c$ & \Pchange \\
Distance (pc) & $\distRSPup \pm \distRSPuperr$ (fixed) \\
  \hline  \noalign{\smallskip}
Radius ($R_\odot$) & \meanRadius\ (\ampRadius) \\
Eff. temperature ($K$) & \meanTeff\ (\ampTeff) \\
Bolom. luminosity ($L_\odot$) & \meanLuminosity\ (\ampLuminosity) \\
Bolometric magnitude & \meanBolometricmag\ (\ampBolometricmag) \\
      \hline
 \end{tabular}
\tablefoot{$^a$ $\theta_0$ is the limb-darkened disk (Rosseland) angular diameter at phase zero, and $\langle \theta \rangle$ the phase-average mean angular diameter over the pulsation cycle.
$^b$ Period at the reference epoch MJD$_0$.
$^c$ Rate of period change as shown in Fig.~\ref{period_var}.
}
\end{table}

\subsection{Projection factor \label{spips-pfactor}}

Considering the complete radial velocity data set, we obtain a projection factor of $p = \pfactorRSPup$ with a statistical uncertainty from the fit of $\sigma_\mathrm{stat} = \pm \pfactorRSPuperrstat$.

The primary source of systematic error on $p$ is the uncertainty on the adopted light echo distance.
As the $p$-factor and the distance are fully degenerate parameters, the $\pm 4.2\%$ distance error bar directly translates into a $\sigma_\mathrm{dist} = \pm 0.053$ uncertainty on $p$.

As shown by \citetads{2014A&A...566L..10A} and \citetads{2016MNRAS.463.1707A}, the cycle-to-cycle repeatability of the velocity curve of long-period Cepheids is imperfect.
\citetads{2016MNRAS.455.4231A} demonstrated that for $\ell$\,Car, variations of the $p$-factor of 5\% are observed between cycles.
To quantify this effect for RS\,Pup, we adjusted distinct SPIPS models on the four cycles sampled by \citetads{2014A&A...566L..10A}.
The results are shown in Figs.~\ref{cycle1} to \ref{cycle4}.
We observe a standard deviation of $\sigma = 0.028$ over the four $p$-factor values derived for the different cycles that we translate into a systematic uncertainty of $\sigma_\mathrm{cycle} = \pm 0.014$ on the $p$-factor.
The SPIPS models resulting from the separate fit of the radial velocity datasets of \citetads{2004A&A...415..531S} and \citetads{2014A&A...566L..10A} are presented in Figs.~\ref{stormvel} and \ref{andersonvel}, respectively.
The derived $p$-factors from these two datasets do not show any significant bias beyond $\sigma_\mathrm{cycle}$.

We assumed in the SPIPS model that the $p$-factor is constant during the pulsation cycle of the star.
This is a simplification, as the $p$-factor is proportional to the limb darkening, which is known to change with the effective temperature of the star.
The amplitude of the $p$-factor variation induced by the changing limb darkening is expected to be small.
The effective temperature of RS\,Pup changes by $1300$\,K during its pulsation ($4600-5900$\,K, Fig.~\ref{RSPup-SPIPS} and Table~\ref{SPIPS-params}).
\citetads{2013A&A...554A..98N} presented predictions of the limb-darkening corrections applicable to interferometric angular diameter measurements based on a spherical implementation of Kurucz's ATLAS models.
For the temperature range of RS\,Pup considering $\log g \approx 1.0$ and $M\approx 10\,M_\odot$), the listed correction factor $k= \theta_\mathrm{UD} / \theta_\mathrm{LD}$ in the $V$ band (in which the spectroscopic measurements are obtained) ranges from $k_V=0.9116$ (4600\,K) to $k_V=0.9161$ (5900\,K) over the cycle.
We consider here that this variation of 0.5\% is negligible compared to the other sources of systematic uncertainty (distance and cycle-to-cycle variations).
%

In summary, combining the systematic uncertainties through $\sigma_\mathrm{syst} = (\sigma_\mathrm{dist}^2 + \sigma_\mathrm{cycle}^2)^{1/2}$ ,
we obtain the $p$-factor of RS\,Pup for the cross-correlation radial velocity method:
\begin{equation}
p = \pfactorRSPup \pm \pfactorRSPuperrstat \pm \pfactorRSPuperrsyst = \pfactorRSPup \pm \pfactorRSPuperrtot\ (\pm \relerrpfactor \%).
\end{equation}

\subsection{Color excess and circumstellar envelope \label{envelope}}
We derive a color excess $E(B-V) = \EBV \pm \EBVerr$, higher than the value obtained by \citetads{2007A&A...476...73F}, who list $E(B-V) = 0.457 \pm 0.009$ for RS\,Pup.
The possible presence of an excess emission in the infrared $K$ ($\lambda \approx 2.2\,\mu$m) and $H$ ($\lambda \approx 1.6\,\mu$m) bands is adjusted as a parameter by the SPIPS code.
For RS\,Pup, we detect a moderately significant excess emission of $\Delta m_K = \excessK \pm \excessKerr$ in the $K$ band, and marginal in the $H$ band ($\Delta m_H = \excessH \pm \excessHerr$\,mag).
This low level of excess emission is in agreement with \citetads{2009A&A...498..425K}, who did not detect a photometric excess in the $K$ band, although a considerable excess flux is found in the thermal infrared ($10\,\mu$m) and at longer wavelengths.
We note that the best-fit infrared excess values for the different pulsation cycles of RS\,Pup (Fig.~\ref{cycle1} to \ref{andersonvel}) are consistent within a few millimagnitudes.

\subsection{Limit on the presence of companions}
\begin{figure}[]
\centering
\includegraphics[width=\hsize]{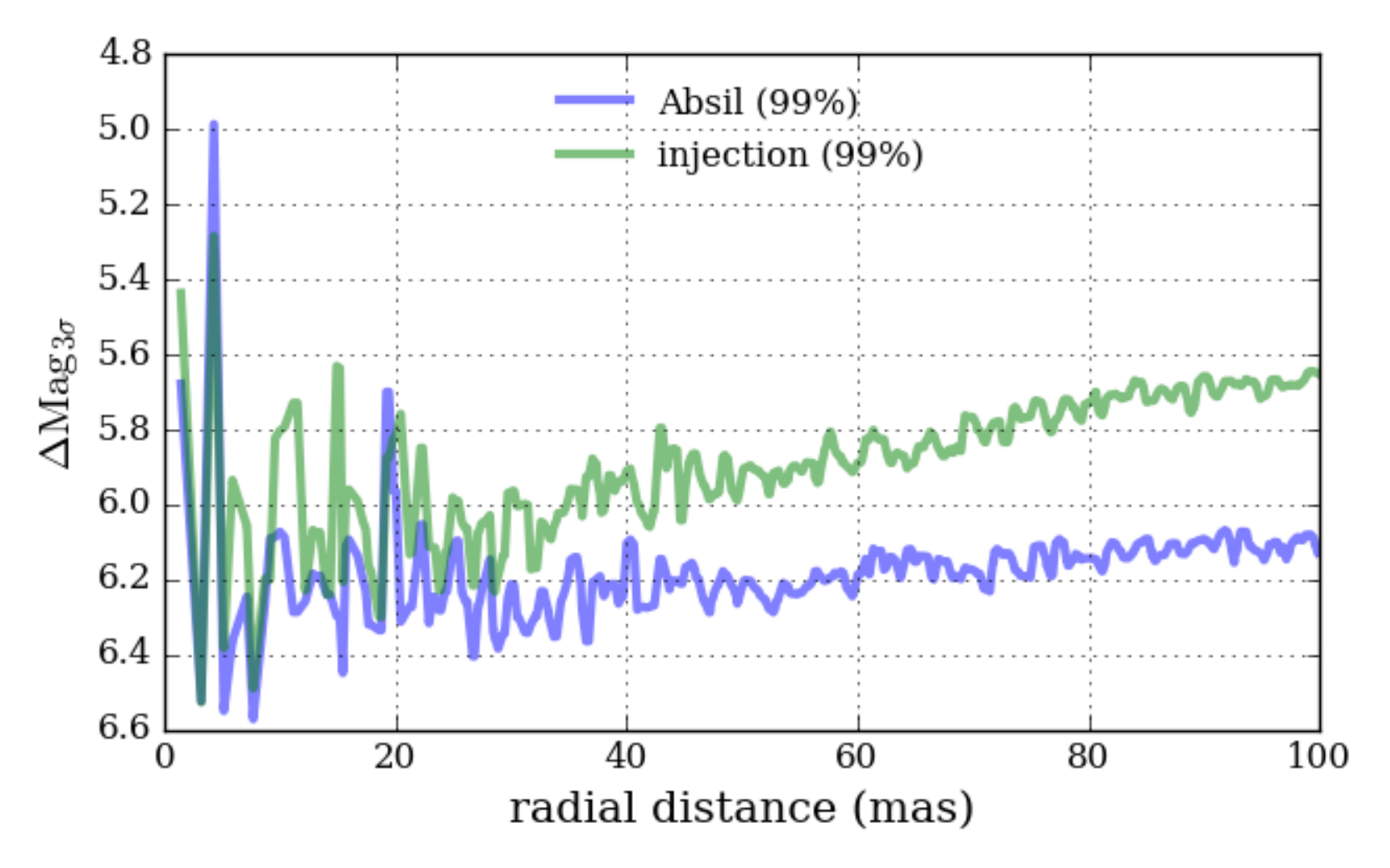}
\includegraphics[width=\hsize]{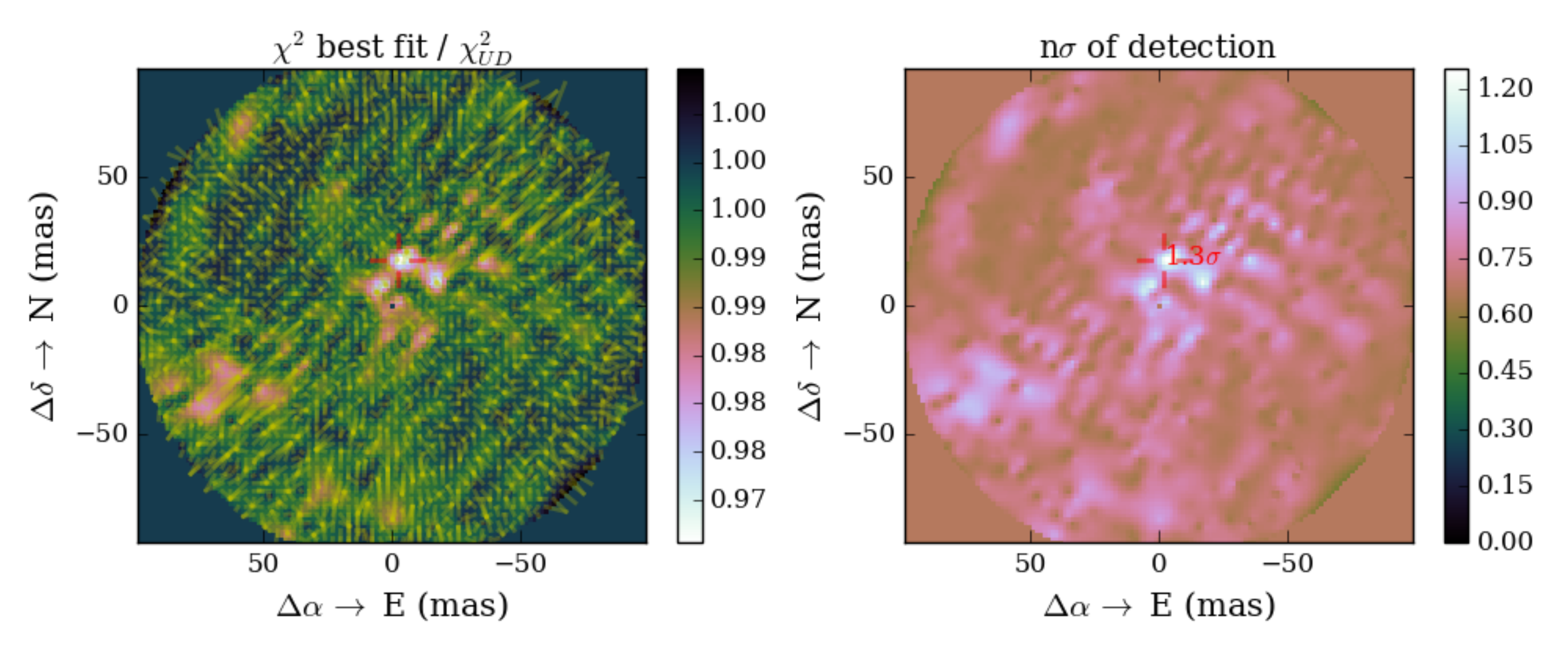}
\caption{{\it Top panel:} Upper limit ($3\sigma$) of the flux contribution of companions of RS\,Pup as a function of the angular separation from the Cepheid.
The limits obtained using the approaches of \citetads{2011A&A...535A..68A} and \citetads{2015A&A...579A..68G} are shown separately. 
{\it Bottom panel:} map of the $\chi^2$ of the best binary model fit (left) and statistical significance of the detection (right). No significant source is found in the field of view.
\label{rspup-candid}}
\end{figure}

We checked for the presence of a companion in the PIONIER interferometric data using the companion analysis and non-detection in interferometric data algorithm (CANDID, \citeads{2015A&A...579A..68G}).
The interferometric observables are particularly sensitive to the presence of companions down to high contrast ratios and small separations, as demonstrated, for instance, by \citetads{2011A&A...535A..68A}, \citetads{2013A&A...552A..21G} and \citetads{2014A&A...561L...3G}.
We did not detect any secondary source, ruling out the presence of a stellar companion with a contrast in the $H$ band less than approximately 6\,magnitudes (flux ratio $f/f_\mathrm{Cepheid} = 0.4\%$) within 40\,mas of the Cepheid (Fig.~\ref{rspup-candid}).

\begin{figure}[]
\centering
\includegraphics[width=\hsize]{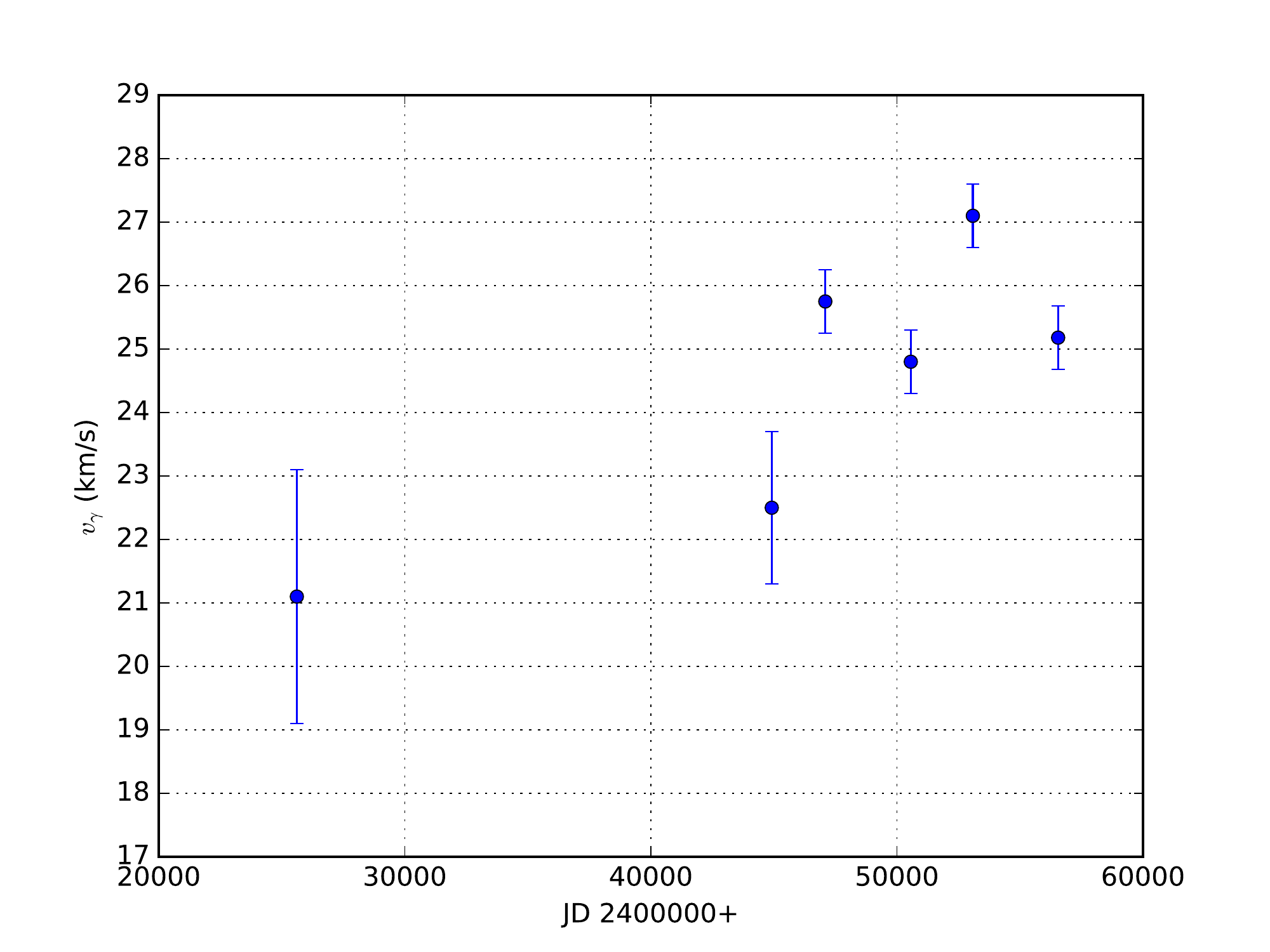} 
\caption{Observed $\gamma$-velocity of RS\,Pup. \label{rsvgamma-new}}
\end{figure}

The $\gamma$-velocity of RS\,Pup measured using the cross-correlation technique is presented in Fig.~\ref{rsvgamma-new}, and the values are listed in Table~\ref{Tabvgamma}.
The cycle-to-cycle random variation of the amplitude of the radial velocity of RS\,Pup \citepads{2014A&A...566L..10A} may induce systematic uncertainties on the determination of the $\gamma$-velocity.
This will particularly be the case if the radial velocity phase coverage is incomplete.
For this reason, while the amplitude of the fluctuations appears significant, it is difficult to conclude that it is caused by a companion.
It is interesting to note that the $\gamma$-velocity value depends on the technique used for the radial velocity measurement: \citetads{2008A&A...489.1255N} find a $\gamma$-velocity of $v_\gamma = -25.7 \pm 0.2$\,km\,s$^{-1}$ for RS\,Pup after correction of the $\gamma$-asymmetry of its spectral lines.
The $\gamma$-velocity can also depend on which lines are included in the cross-correlation mask.

\begin{table}[!]
\caption{$\gamma$-velocities of RS\,Pup from the cross-correlation technique.}
\label{Tabvgamma}
\begin{center}
\begin{tabular}{lcl}
\hline
\hline \noalign{\smallskip}
Mean JD & $v_{\gamma}$ (km\,s$^{-1}$) & Reference\\
\noalign{\smallskip}
\hline \noalign{\smallskip}
2\,425\,612 & $21.1 \pm 2.0$ & {\citetads{1939ApJ....89..356J}}\\
2\,444\,913 & $22.5 \pm 1.2$ & {\citetads{1988ApJS...66...43B}}\\
2\,447\,090 & $25.8 \pm 0.5$ & {\citetads{2004A&A...415..531S}} \\
2\,450\,563 & $24.8 \pm 0.5$ & {\citetads{2002ApJS..140..465B}} \\
2\,453\,085 & $27.1 \pm 0.5$ & {\citetads{2009A&A...502..951N}} \\
2\,456\,551 & $25.2 \pm 0.5$ & {\citetads{2014A&A...566L..10A}} \\
\hline
\end{tabular}
\end{center}
\end{table}

\section{Discussion\label{discussion}}

\begin{table}
 \caption{Measured (top section) $p$-factor values of RS\,Pup and $\ell$\,Car and predictions from period-$p$-factor relations (bottom section).}
 \label{table-pfactors}
 \centering
 \renewcommand{\arraystretch}{1.2}
 \begin{tabular}{lcc}
  \hline
  \hline
  Reference    & RS\,Pup & $\ell$\,Car \\ 
  \hline
  {\citetads{2016A&A...587A.117B}} & $-$ & $1.23 \pm 0.12$ \\
  {\citetads{2016MNRAS.455.4231A}}$^c$ & $-$ & $\pfactorLCar \pm \pfactorLCarerr$ \\
  Present work & $\pfactorRSPup \pm \pfactorRSPuperrtot$ & $-$ \\
  \hline 
  {\citetads{1982A&A...109..258B}} & 1.36 & 1.36 \\
  {\citetads{1986PASP...98..881H}} & 1.341 & 1.343 \\
  {\citetads{2005ApJ...627..224G}} & $1.337 \pm 0.038$ & $1.347 \pm 0.037$ \\
  {\citetads{2007A&A...474..975G}}$^a$ & $1.270 \pm 0.050$ & $1.270 \pm 0.050$ \\
  {\citetads{2009AIPC.1170...93L}} & $1.196 \pm 0.038$ & $1.201 \pm 0.036$ \\
  {\citetads{2011A&A...534A..94S}} & $1.249 \pm 0.105$ & $1.262 \pm 0.101$ \\
  {\citetads{Neilson:2012qy}} & $1.140 \pm 0.003$ & $1.146 \pm 0.003$ \\
  {\citetads{2013A&A...550A..70G}} & $1.112 \pm 0.030$  & $1.128 \pm 0.030$  \\
  {\citetads{Nardetto:2014kx}} & $1.181 \pm 0.019$ & $1.186 \pm 0.018$ \\
  \hline
 \end{tabular}
 \tablefoot{$^a$ Constant $p$-factor value; $^b$ excluding FF\,Aql; $^c$ average of the measured values.}
\end{table}

For a review of the current open questions related to the $p$-factor, in particular in the context of the interferometric version of the BW technique, we refer to \citetads{2009AIPC.1170....3B, 2012JAVSO..40..256B}.

A summary of the available predictions and measurements of the $p$-factors of RS\,Pup and of the similar long-period Cepheid $\ell$\,Car is presented in Table~\ref{table-pfactors}.
Most authors based their BW distance determination on the linear period-$p$-factor relation established by \citetads{1986PASP...98..881H, 1989ApJ...341.1004H}: $p=1.39 - 0.03 \log P$. Owing to the weak dependence on period, the $p$-factors predicted for RS\,Pup and $\ell$\,Car by this relation are both very close to $p=1.34$.
The theoretical calibration of the period-$p$-factor (P$p$) relation by \citetads{Neilson:2012qy} gives a geometric $p$-factor of $p_0 = [1.402 \pm 0.002] - [0.0440 \pm 0.0015] \log P$ ($V$ band, spherical model, linear law), to be multiplied by the period-dependent velocity gradient and differential velocity corrections introduced by \citetads{2007A&A...471..661N}.
%
The recent work by \citetads{Nardetto:2014kx} including $\delta$~Scuti stars confirms the P$p$ relation by \citetads{2009A&A...502..951N} and proposes a common P$p$ relation between Cepheids and $\delta$~Scuti stars ($p = [1.31 \pm 0.01] - [0.08 \pm 0.01] \log P$).
The \citetads{Nardetto:2014kx} relation yields $p=1.181$ for RS\,Pup and $p=1.186$ for $\ell$\,Car.
The relation from \citetads{2011A&A...534A..94S} is much steeper ($p = [1.550 \pm 0.04] - [0.186 \pm 0.06] \log P$).
\citetads{2007A&A...474..975G} used the Cepheid trigonometric parallaxes from \citetads{2007AJ....133.1810B} to derive a P$p$ relation of the form $p = [1.28 \pm 0.15] - [0.01 \pm 0.16] \log P$, which is consistent with a constant $p$-factor with $p = 1.27 \pm 0.05$.

\begin{figure}[]
\centering
\includegraphics[width=\hsize]{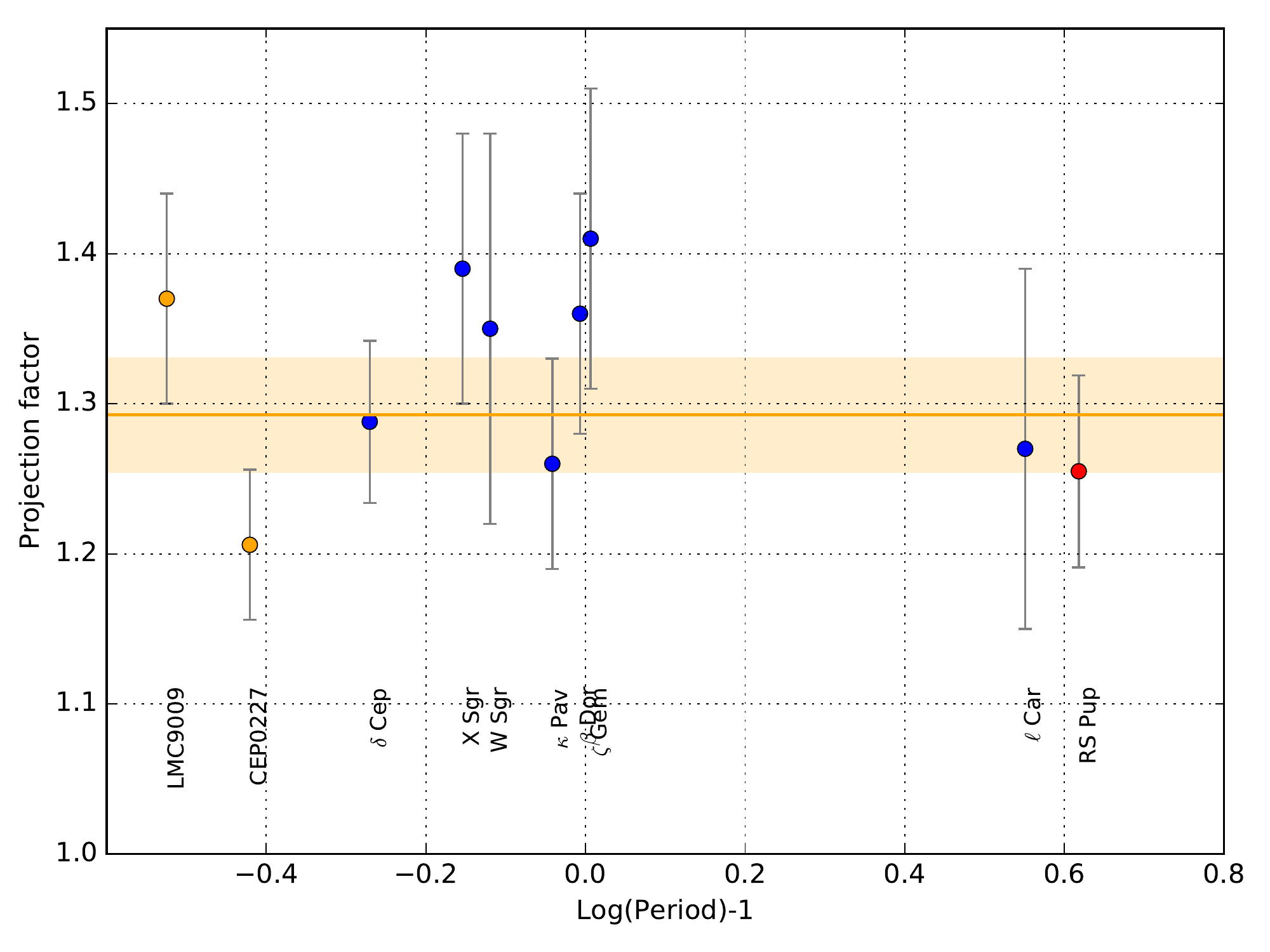} 
\caption{Distribution of the measured $p$-factors of Cepheids with better than 10\% relative accuracy.
The references for the different measurements (except RS\,Pup) are listed in \citetads{2016A&A...587A.117B}.
The blue points use HST-FGS distances \citepads{2002AJ....124.1695B,2007AJ....133.1810B}, the orange points are the LMC eclipsing Cepheids \citepads{2013MNRAS.436..953P,2015ApJ...815...28G}.
The solid line and orange shaded area represent the weighted average $\overline{p}=\meanp \pm \meanperr$. \label{Ppfactor}}
\end{figure}

Figure~\ref{Ppfactor} gives an overview of the available measurements of $p$-factors of Cepheids, including the Type II Cepheid $\kappa$\,Pav \citepads{2015A&A...576A..64B}.
We selected for this plot the $p$-factor values with a relative accuracy better than 10\%.
We removed from the sample the binary Cepheid FF\,Aql for which the HST/FGS distance is questionable (see the discussion, e.g., in \citeads{2016A&A...587A.117B} and \citeads{2013ApJ...772L..10T}).
As shown by \citetads{2016ApJS..226...18A}, the presence of a companion can bias the parallax.
The weighted average of the selected measurements is $\overline{p}=\meanp \pm \meanperr$, and the reduced $\chi^2$ of the measurements with respect to this constant value is $\chi^2_\mathrm{red} = \pchired$.
If we include FF Aql in the sample, we obtain $\overline{p} = \meanpwithFFAql$.
The uncertainty of $\overline{p}$ was computed from the combination of the error bars of the independent measurements of OGLE-LMC-CEP-0227 ($P=3.80$\,d, $p=1.21 \pm 0.05$, \citeads{2013MNRAS.436..953P}), $\delta$\,Cep ($P=5.37$\,d, $p=1.288 \pm 0.054$, \citeads{2015A&A...584A..80M}), and the present measurement of RS\,Pup ($P=41.5$\,d, $p=\pfactorRSPup \pm \pfactorRSPuperrtot$).
We did not average the error bars of the different $p$-factor measurements from the HST/FGS distances as the degree of correlation between them and the possible associated systematics are uncertain.
For the same reason, we did not average the uncertainties of the two $p$-factor measurements of binary Cepheids in the LMC from \citetads{2013MNRAS.436..953P} and \citetads{2015ApJ...815...28G}, and we selected only the best $p$-factor of $\delta$\,Cep derived by \citetads{2015A&A...584A..80M}.
In agreement with the present results, \citetads{2016A&A...587A.117B} also concluded from a fit to the complete sample of measured $p$-factors that a constant value of $\overline{p} = 1.324 \pm 0.024$ ($1\sigma$ from our value) reproduces the measurements.

The good agreement of the constant $p$-factor model $\overline{p}=\meanp \pm \meanperr$ with the measurements indicates that this coefficient is  mildly variable over a broad range of Cepheid periods (3.0 to 41.5\,days).
This result can be explained by the relatively narrow range of effective temperature and gravity of Cepheids, which results in a minor variation of their limb darkening.
\citetads{2013A&A...554A..98N} predict changes of the limb-darkening coefficient $k = \theta_\mathrm{UD}/\theta_\mathrm{LD}$ of only
a few percent in the $V$ band over the full range of classical Cepheid properties.
The difference is even smaller at longer wavelengths.
The spherical models in the $V$ band by these authors give $k = 0.9337$ for the hottest phase of a short-period Cepheid (7000\,K, $\log g = 2.0$, $m = 5\,M_\odot$), less than 2.5\% away from the value $k = 0.9116$ obtained for the coolest phase of RS\,Pup (4600\,K, $\log g = 1.0$, $m=10\,M_\odot$).
The mild dependence of the $p$-factor on the period is consistent with the P$p$ relation proposed by \citetads{2007A&A...474..975G}.

The precision of the parallaxes of the first data release of \emph{Gaia}-TGAS \citepads{2016A&A...595A...4L} is too low to
accurately determine the $p$-factor of nearby Cepheids (see, e.g., \citeads{2016arXiv160905175C}).
The availability in 2017 of the second \emph{Gaia} data release \citepads{2016A&A...595A...1G} will provide very accurate parallaxes for hundreds of Galactic Cepheids, however, including RS\,Pup,
which will be among the longest periods in the sample. 
The ongoing observations of a sample of 18 long-period Cepheids by \citetads{2016ApJ...825...11C} using the spatial scanning technique with the HST/WFC3 has started to provide accurate parallaxes with accuracies of $\pm 30\,\mu$s for these rare pulsators.
At a later stage, accurate broadband epoch photometry will also be included in the \emph{Gaia} data releases (see, e.g., \citeads{2016A&A...595A.133C}).
Combining \emph{Gaia} data with archival observations, the SPIPS technique will enable a very accurate calibration of the P$p$ relation of Cepheids, and therefore of their distance scale, which is still today an essential ingredient in determining the local value of $H_0$ \citepads{2016ApJ...826...56R}.
For the nearest Cepheids of the \emph{Gaia} and HST/WFC3 samples, the availability of interferometric angular diameters will significantly improve the quality of the determination of their parameters thanks to the resolution of the usual degeneracy between effective temperature and interstellar reddening.
Even for distant Cepheids, however, whose angular diameters cannot be measured directly, the robustness of the SPIPS algorithm will enable an accurate calibration of their physical properties, including the $p$-factor, once their parallaxes are known.


%

\begin{acknowledgements}
We would like to thank Vello Tabur and Stanley Walker for sending us their unpublished photometric observations, and Marcella Marconi.
We acknowledge with thanks the variable star observations from the AAVSO International Database contributed by observers worldwide and used in this research.
The authors acknowledge the support of the French Agence Nationale de la Recherche (ANR), under grant ANR-15-CE31-0012-01 (project UnlockCepheids).
PK, AG, and WG acknowledge support of the French-Chilean exchange program ECOS-Sud/CONICYT (C13U01).
W.G. and G.P. gratefully acknowledge financial support for this work from the BASAL Centro de Astrofisica y Tecnologias Afines (CATA) PFB-06/2007. 
W.G. also acknowledges financial support from the Millenium Institute of Astrophysics (MAS) of the Iniciativa Cientifica Milenio del Ministerio de Economia, Fomento y Turismo de Chile, project IC120009.
We acknowledge financial support from the ``Programme National de Physique Stellaire" (PNPS) of CNRS/INSU, France.
LSz acknowledges support from the ESTEC Contract No.\,4000106398/12/NL/KML.
The research leading to these results  has received funding from the European Research Council (ERC) under the European Union's Horizon 2020 research and innovation programme (grant agreement No 695099).
This research made use of Astropy\footnote{Available at \url{http://www.astropy.org/}}, a community-developed core Python package for Astronomy \citepads{2013A&A...558A..33A}.
This work has made use of data from the European Space Agency (ESA) mission {\it Gaia} (\url{http://www.cosmos.esa.int/gaia}), processed by the {\it Gaia} Data Processing and Analysis Consortium (DPAC, \url{http://www.cosmos.esa.int/web/gaia/dpac/consortium}). Funding for the DPAC has been provided by national institutions, in particular the institutions participating in the {\it Gaia} Multilateral Agreement.
We used the SIMBAD and VIZIER databases at the CDS, Strasbourg (France), and NASA's Astrophysics Data System Bibliographic Services.
\end{acknowledgements}

\bibliographystyle{aa} 
\bibliography{biblioRSPup4}

\newpage

\begin{appendix}

\section{Photometric measurements}

\input{RSPupTables}

\section{SPIPS analysis of separate pulsation cycles}

\begin{figure*}[]
\centering
\includegraphics[width=15cm]{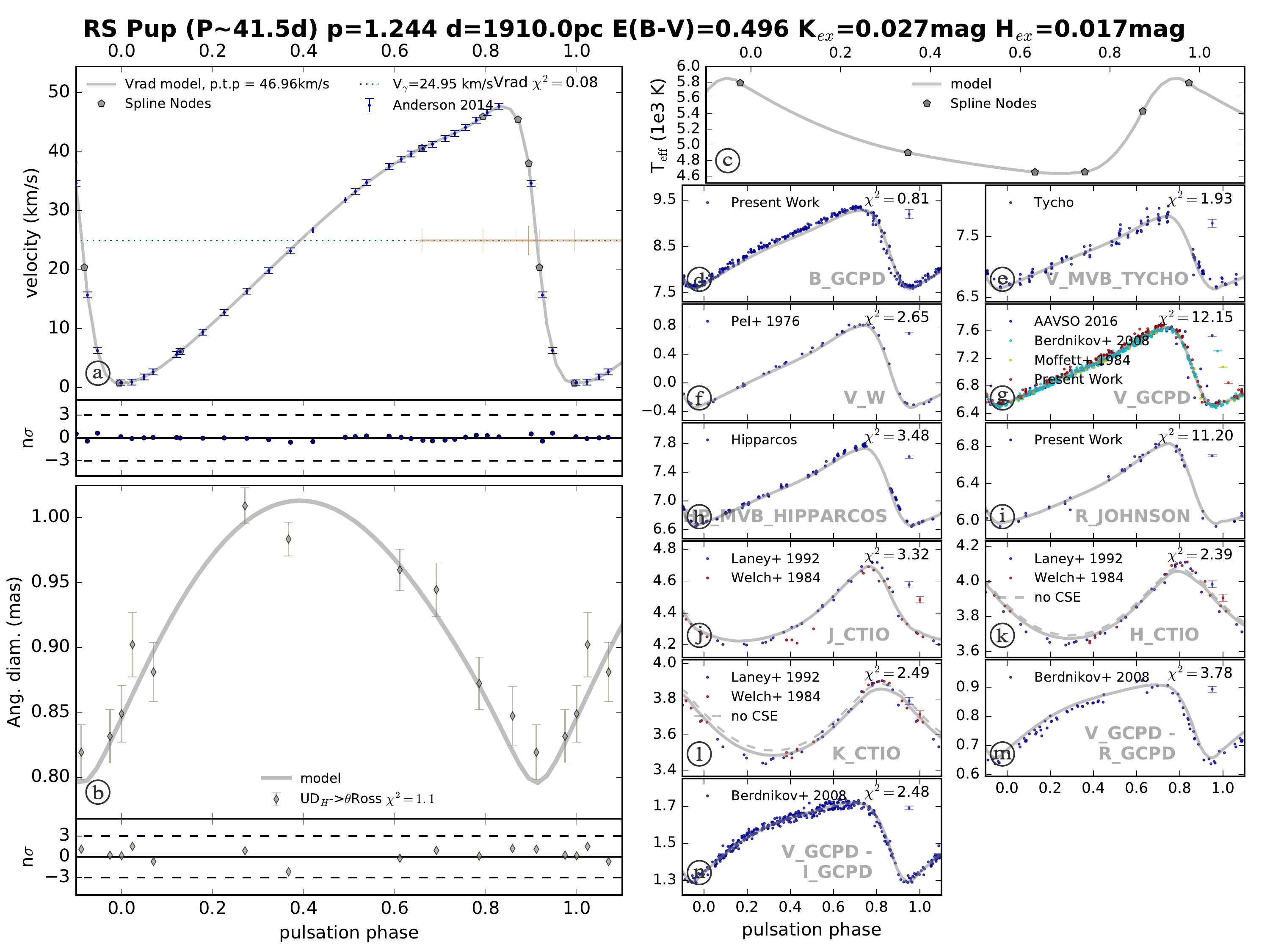} 
\caption{SPIPS model of RS\,Pup for the radial velocities of Cycle 1 of \citetads{2014A&A...566L..10A}. \label{cycle1}}
\end{figure*}

\begin{figure*}[]
\centering
\includegraphics[width=15cm]{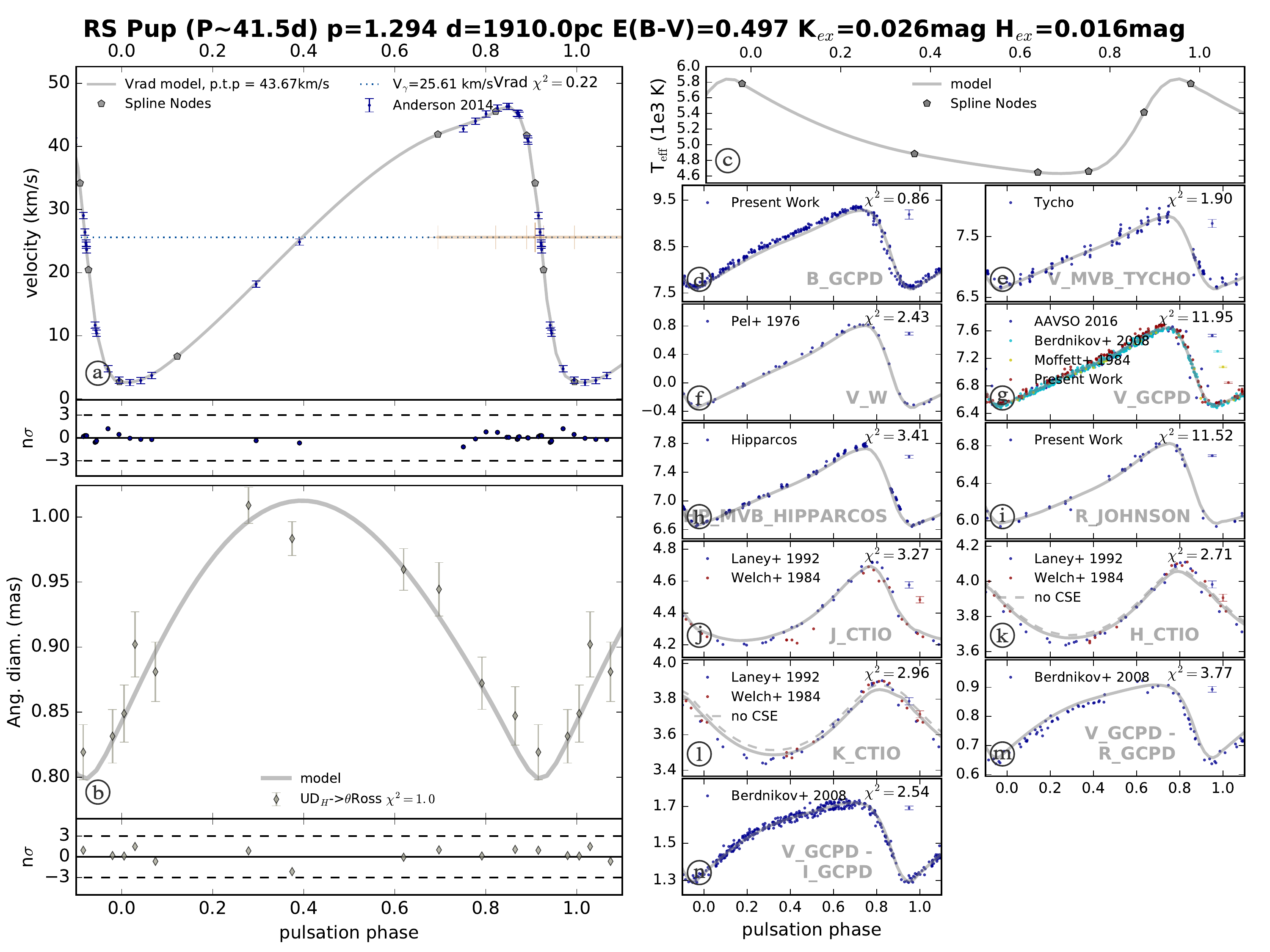} 
\caption{Same as Fig.~\ref{cycle1} for Cycle 2. \label{cycle2}}
\end{figure*}

\begin{figure*}[]
\centering
\includegraphics[width=15cm]{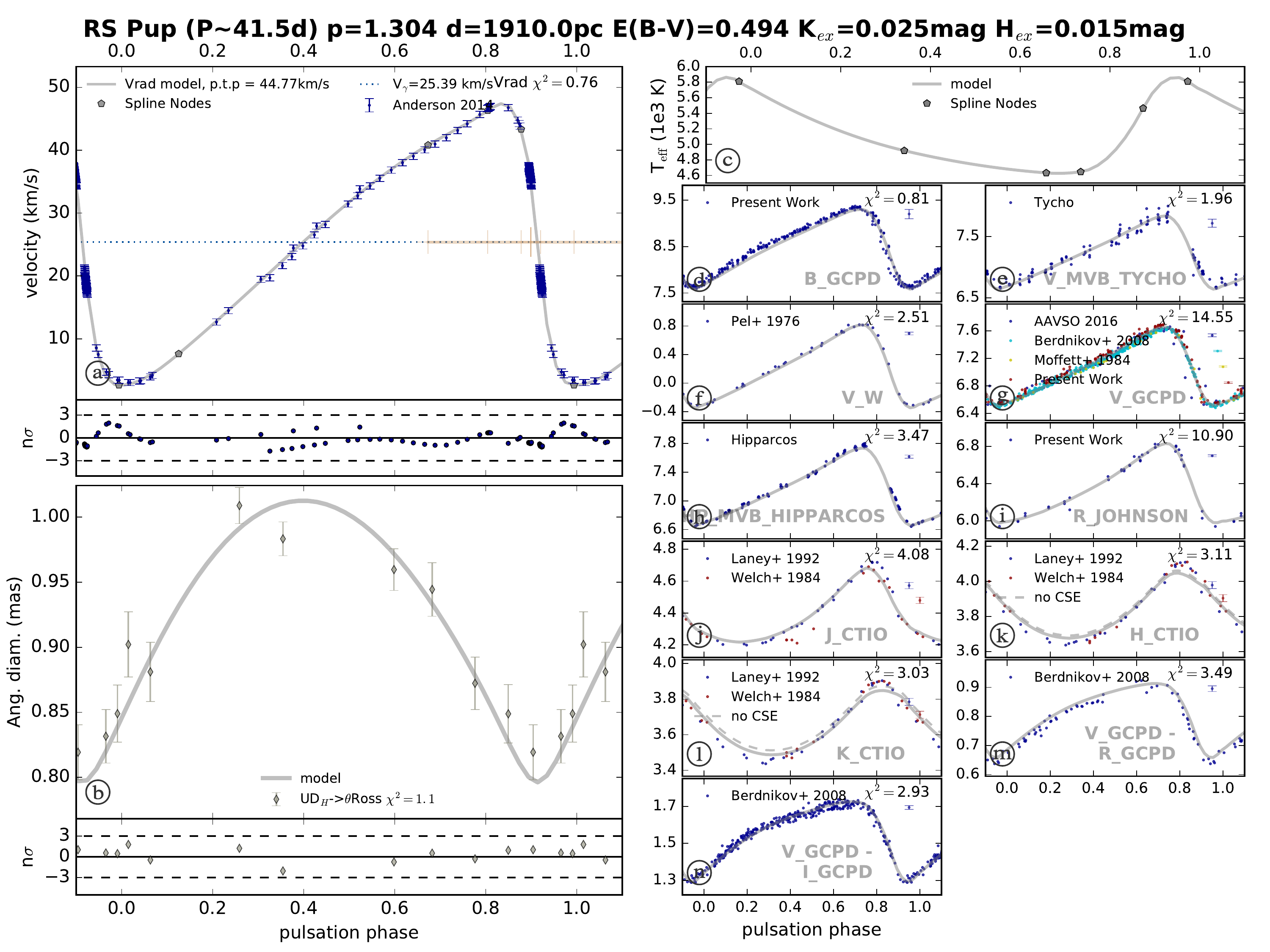} 
\caption{Same as Fig.~\ref{cycle1} for Cycle 3. \label{cycle3}}
\end{figure*}

\begin{figure*}[]
\centering
\includegraphics[width=15cm]{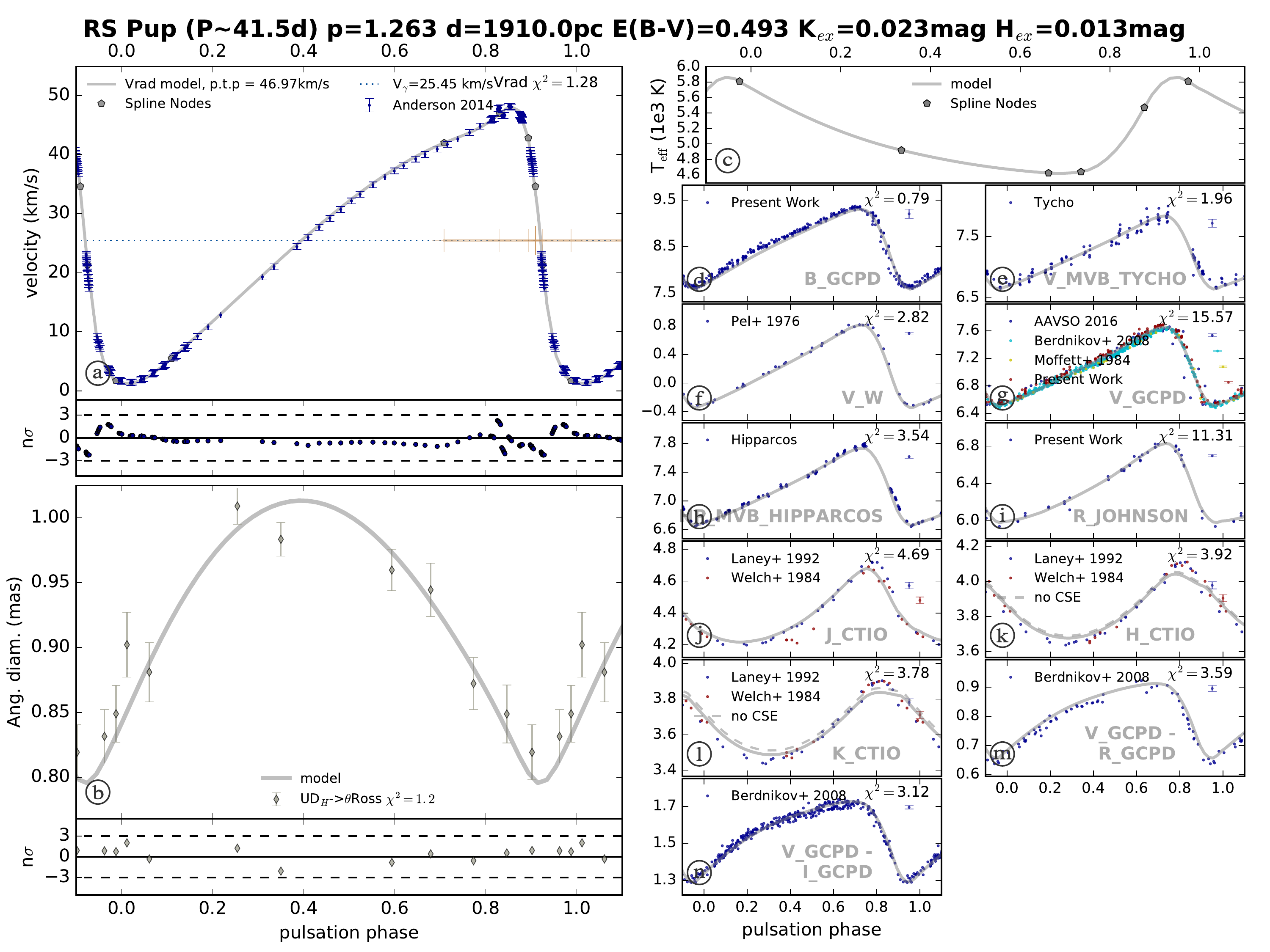} 
\caption{Same as Fig.~\ref{cycle1} for Cycle 4. \label{cycle4}}
\end{figure*}

\begin{figure*}[]
\centering
\includegraphics[width=15cm]{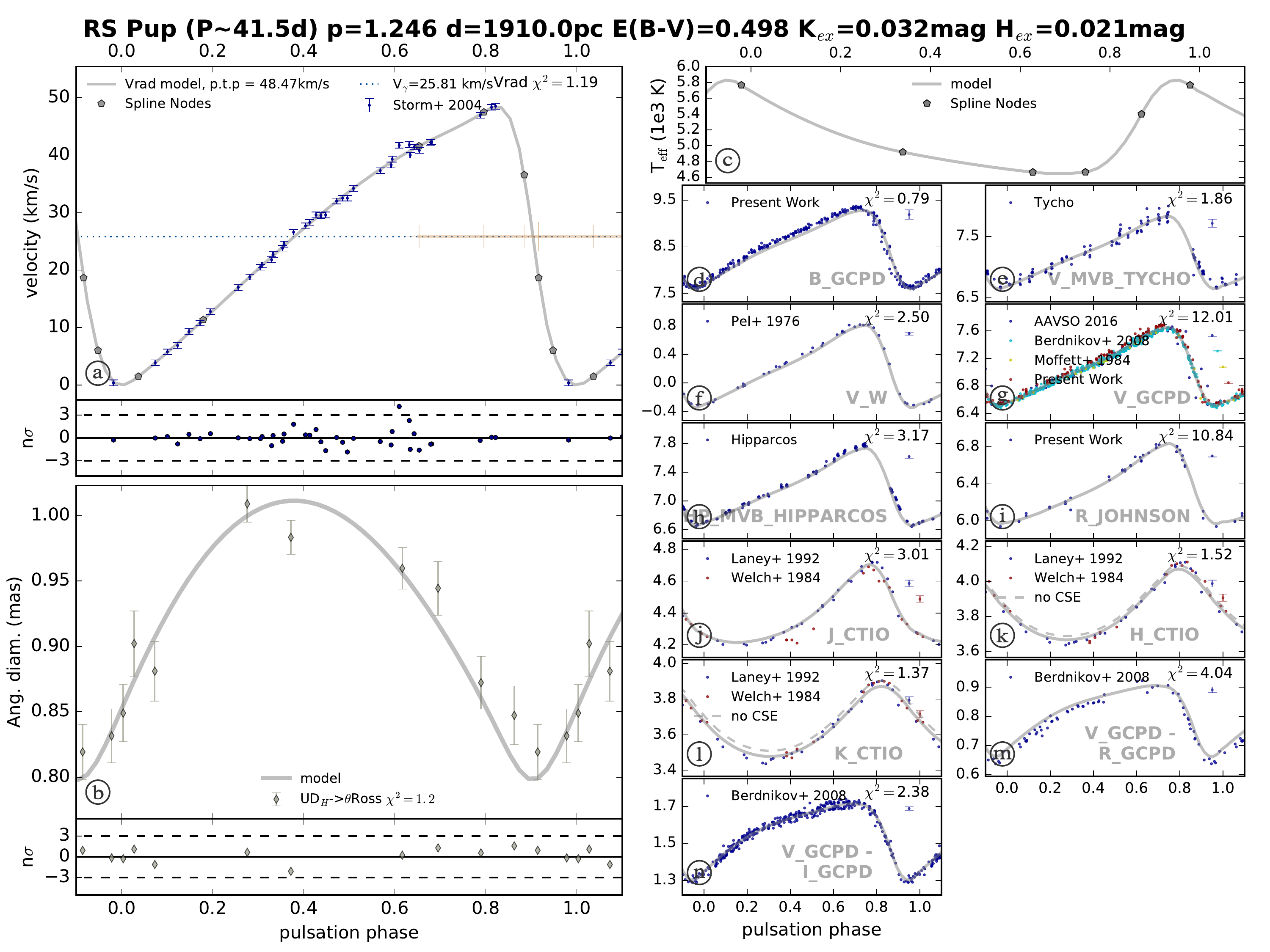} 
\caption{SPIPS model of RS\,Pup for the radial velocities collected exclusively by \citetads{2004A&A...415..531S}. \label{stormvel}}
\end{figure*}

\begin{figure*}[]
\centering
\includegraphics[width=15cm]{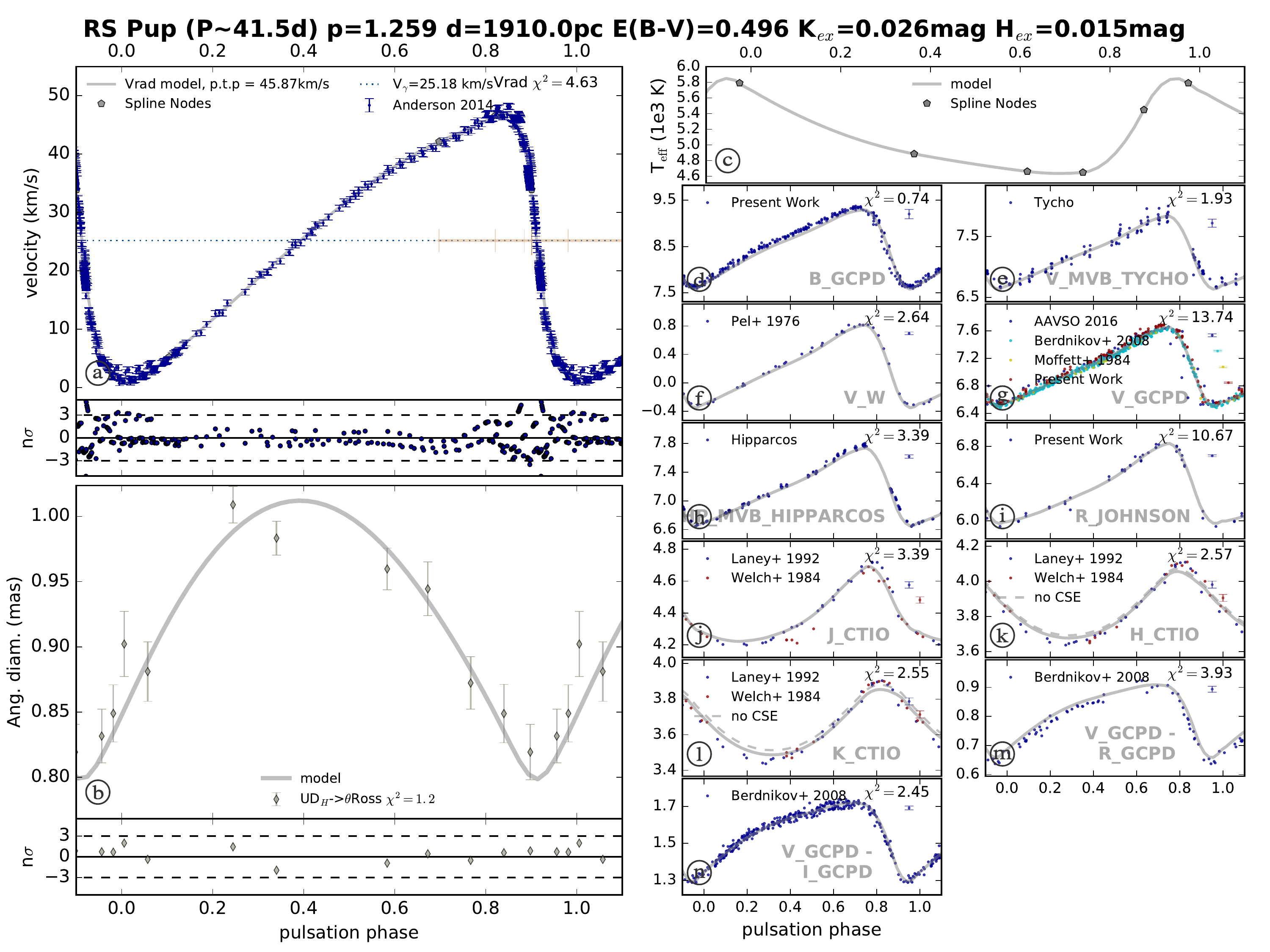} 
\caption{Same as Fig.~\ref{stormvel} for a combination of all cycles of the radial velocity observations by \citetads{2014A&A...566L..10A}. \label{andersonvel}}
\end{figure*}

We present here the results of the SPIPS modeling of the four pulsation cycles of RS\,Pup observed by \citetads{2014A&A...566L..10A}.
We keep in the dataset only the radial velocity data of one cycle, while keeping all the other datasets unchanged.
The results are presented in Fig.~\ref{cycle1} to \ref{cycle4}.
We also present in Fig.~\ref{stormvel} and \ref{andersonvel} the best-fit SPIPS solutions obtained considering separately the radial velocity datasets of \citetads{2004A&A...415..531S} and \citetads{2014A&A...566L..10A}, respectively.

\end{appendix}

\end{document}

%% file: RSPupTables.tex
\twocolumn
\begin{table}
\caption{$O-C$ residuals for RS\,Puppis.
JD$_\odot$ is the heliocentric moment of maximum brightness.
$E$ is the epoch number as calculated from the ephemeris $C = 2455501.254 + 41.49\times E$.
$W$ is the weight assigned to the $O-C$ value (1, 2, or 3 depending on the quality of the light curve).
The $^*$ symbol in the references indicates unpublished data.
\label{tab-oc}}
\end{table}
\tablehead{\hline \hline
JD$_\odot$ & E\ \  &   $O-C$ &  $W$ &  Reference \\
$-2.4 \times 10^6$  &    &    [d] & & \\
\hline \noalign{\smallskip}}
\tabletail{\hline}
\begin{supertabular}{crrcl}
15050.6  &  $-$977 & 85.1 &  1 & \citetads{1903AnCap...9....1I} \\
15424.6  &  $-$968 & 85.7 &  1 & \citetads{1927BHarO.848R..14G}\\
16830.1  &  $-$934 & 80.5 &  1 & \citetads{1927BHarO.848R..14G} \\
18277.9  &  $-$899 & 76.2 &  1 & \citetads{1927BHarO.848R..14G} \\
19723.4  &  $-$864 & 69.5 &  1 & \citetads{1927BHarO.848R..14G} \\
20963.2  &  $-$834 & 64.6 &  1 & \citetads{1927BHarO.848R..14G} \\
22824.1  &  $-$789 & 58.5 &  1 & \citetads{1927BHarO.848R..14G} \\
26006.61 &  $-$712 & 46.24  & 1 & \citetads{1939AnBos...8...31V} \\
26379.42 &  $-$703 & 45.64  & 1 & \citetads{1939AnBos...8...31V} \\
26752.55 &  $-$694 & 45.36 & 1 & \citetads{1939AnBos...8...31V} \\
27042.15 &  $-$687 & 44.53 & 1 & \citetads{1939AnBos...8...31V} \\
27456.32 &  $-$677 & 43.80  & 1 & \citetads{1939AnBos...8...31V} \\
27828.94 &  $-$668 & 43.01 & 1 & \citetads{1939AnBos...8...31V} \\
28201.64 &  $-$659 & 42.30 & 1 & \citetads{1939AnBos...8...31V} \\
34451.700 & $-$508 & 27.366 & 1 & \citetads{1957MNRAS.117..406E}\\
34533.607 & $-$506 & 26.293 & 2 & \citetads{1958BAN....14...81W}\\
34864.914 & $-$498 & 25.680 & 2 & \citetads{1957MNRAS.117..406E}\\
35196.511 & $-$490 & 25.357 & 1 & \citetads{1961ApJS....6..253I} \\
37513.865 & $-$434 & 19.271 & 3 & \citetads{1963MNRAS.127...71W} \\
37638.297 & $-$431 & 19.233 & 3 & \citetads{1964BOTT....3..153M} \\
40701.394 & $-$357 & 12.070 & 3 & \citetads{1976AAS...24..413P} \\
40949.305 & $-$351 & 11.041 & 3 & \citetads{1976AAS...24..413P} \\
41405.313 & $-$340 & 10.659 & 1 & \citetads{1977MmRAS..83...69D}\\
41735.214 & $-$332 &  8.640 & 2 & \citetads{1975ApJS...29..219M}\\
41776.643 & $-$331 &  8.579 & 3 & \citetads{1977MmRAS..83...69D}\\
42065.736 & $-$324 &  7.242 & 3 & \citetads{1977MmRAS..83...69D}\\
42438.472 & $-$315 &  6.568 & 2 & \citetads{1977MmRAS..83...69D}\\
44055.085 & $-$276 &  5.071 & 2 & \citetads{1980PhDT.........9H} \\
44428.140 & $-$267 &  4.716 & 1 & \citetads{1984ApJS...55..389M} \\
44967.089 & $-$254 &  4.295 & 3 & \citetads{1984ApJS...55..389M} \\
47951.011 & $-$182 &  0.937 & 3 & \citetads{1997ESASP1200.....E} \\
48240.973 & $-$175 &  0.469 & 3 & \citetads{1997ESASP1200.....E} \\
48448.945 & $-$170 &  0.991 & 3 & \citetads{1997ESASP1200.....E} \\
48739.114 & $-$163 &  0.730 & 3 & \citetads{1997ESASP1200.....E} \\
49111.661 & $-$154 & $-$0.133 & 1 & AAVSO \\
49360.599 & $-$148 & $-$0.135 & 1 & Walker \& Williams$^*$ \\
49527.027 & $-$144 &  0.333 & 1 & AAVSO\\
49817.467 & $-$137 &  0.343 & 3 & \citetads{1995ASPC...83..349B, 2008yCat.2285....0B} \\
50108.216 & $-$130 &  0.662 & 3 & \citetads{2002ApJS..140..465B} \\
50357.203 & $-$124 &  0.709 & 3 & \citetads{1995ASPC...83..349B, 2008yCat.2285....0B} \\
50522.961 & $-$120 &  0.507 & 3 & \citetads{2002ApJS..140..465B} \\
50564.680 & $-$119 &  0.736 & 1 & \citetads{1995ASPC...83..349B, 2008yCat.2285....0B}\\
51270.917 & $-$102 &  1.643 & 1 & \citetads{1995ASPC...83..349B, 2008yCat.2285....0B}\\
51644.316 &  $-$93 &  1.632 & 2 & \citetads{1995ASPC...83..349B, 2008yCat.2285....0B}\\
51976.659 &  $-$85 &  2.055 & 3 & \citetads{1995ASPC...83..349B, 2008yCat.2285....0B}\\
52350.347 &  $-$76 &  2.333 & 3 & \citetads{1995ASPC...83..349B, 2008yCat.2285....0B}\\
52640.475 &  $-$69 &  2.031 & 3 & \citetads{1995ASPC...83..349B, 2008yCat.2285....0B}\\
52681.984 &  $-$68 &  2.050 & 3 & \citetads{2002AcA....52..397P}\\
53014.205 &  $-$60 &  2.351 & 2 & \citetads{1995ASPC...83..349B, 2008yCat.2285....0B}\\
53055.383 &  $-$59 &  2.039 & 3 & \citetads{2002AcA....52..397P}\\
53096.193 &  $-$58 &  1.359 & 3 & \citetads{1995ASPC...83..349B, 2008yCat.2285....0B}\\
53551.918 &  $-$47 &  0.694 & 2 & Tabur$^*$ \\
53842.397 &  $-$40 &  0.743 & 2 & AAVSO\\
53842.461 &  $-$40 &  0.807 & 3 & Tabur$^*$ \\
53842.495 &  $-$40 &  0.841 & 2 & \citetads{2002AcA....52..397P}\\
54091.419 &  $-$34 &  0.825 & 3 & present work \\
54132.197 & $-$33 & 0.113 & 3 & AAVSO\\
54215.229 & $-$31 & 0.165 & 3 & Tabur$^*$ \\
54215.264 & $-$31 & 0.200 & 3 & \citetads{2002AcA....52..397P} \\
54505.241 & $-$24 & $-$0.253 & 3 & \citetads{2002AcA....52..397P} \\
54546.574 & $-$23 & $-$0.410 & 3 & Tabur$^*$ \\
54546.611 & $-$23 & $-$0.373 & 3 & AAVSO \\
54588.094 & $-$22 & $-$0.380 & 2 & present work \\
54837.503 & $-$16 & 0.089 & 3 & \citetads{2002AcA....52..397P} \\
54878.846 & $-$15 & $-$0.058 & 2 & AAVSO\\
55127.522 &  $-$9 & $-$0.322 & 1 & \citetads{2002AcA....52..397P} \\
55252.067 &  $-$6 & $-$0.247 & 2 & AAVSO\\
55252.640 & $-6$ & $-0.241$ &  3 & \citetads{Kervella:2014lr} \\
55293.713 &  $-$5 & $-$0.091 & 2 & present work \\
55335.074 &  $-$4 & $-$0.220 & 3 & present work \\
55501.254 &  0 & 0.000 & 3 & present work \\
55625.868 &  3 & 0.144 & 2 & AAVSO\\
56041.317 & 13 & 0.693 & 3 & AAVSO\\
56373.179 & 21 & 0.635 & 3 & AAVSO\\
\hline
\end{supertabular}

\twocolumn
\begin{table}
\caption{SMARTS photometry of RS\,Pup. The phases were computed assuming $P = 41.5113$\,days and the maximum light epoch $\mathrm{JD}_\odot = 2455501.254$. \label{tab-smarts}}
\end{table}
\tablehead{\hline \hline \noalign{\smallskip} $\mathrm{JD}_\odot -2.4 \times 10^6$ & Phase & $B$ & $V$ & $R$ \\ \hline \noalign{\smallskip}}
\tabletail{\hline}
\begin{supertabular}{ccccc}
54525.7016 & 0.0000 & 8.907 & 7.217 & $-$ \\
54532.6658 & 0.1678 & 9.211 & 7.559 & 6.690 \\
54536.6723 & 0.2643 & 9.389 & 7.647 & 6.823 \\
54539.6256 & 0.3354 & 9.257 & $-$ & $-$ \\
54543.7118 & 0.4339 & 8.094 & $-$ & $-$ \\
54558.5980 & 0.7925 & 8.508 & $-$ & $-$ \\
54563.5447 & 0.9116 & 8.732 & $-$ & $-$ \\
54567.6262 & 0.0100 & 8.936 & 7.255 & $-$ \\
54569.5540 & 0.0564 & 9.061 & 7.304 & $-$ \\
54576.6180 & 0.2266 & 9.370 & 7.632 & $-$ \\
54578.5235 & 0.2725 & 9.418 & 7.657 & $-$ \\
54585.5011 & 0.4406 & 7.999 & $-$ & $-$ \\
54587.5767 & 0.4906 & 7.717 & 6.565 & $-$ \\
54592.5966 & 0.6115 & 7.979 & $-$ & $-$ \\
54596.4998 & 0.7055 & 8.280 & $-$ & $-$ \\
54606.4899 & 0.9462 & 8.778 & $-$ & $-$ \\
54608.4538 & 0.9935 & 8.897 & $-$ & $-$ \\
54610.5160 & 0.0432 & 8.999 & 7.319 & 6.476 \\
54616.5077 & 0.1875 & 9.226 & 7.626 & 6.684 \\
54623.4738 & 0.3553 & 9.046 & 7.502 & 6.745 \\
54628.4718 & 0.4757 & 7.759 & 6.562 & $-$ \\
54635.4700 & 0.6443 & 8.021 & 6.743 & $-$ \\
54643.4491 & 0.8365 & 8.651 & 7.045 & $-$ \\
54725.8512 & 0.8216 & 8.581 & $-$ & $-$ \\
54729.8645 & 0.9182 & 8.728 & $-$ & $-$ \\
54733.8631 & 0.0146 & 8.901 & $-$ & $-$ \\
54737.8615 & 0.1109 & 9.158 & $-$ & $-$ \\
54741.8534 & 0.2071 & 9.369 & $-$ & $-$ \\
54745.8538 & 0.3034 & 9.358 & $-$ & $-$ \\
54749.8616 & 0.4000 & 8.516 & $-$ & $-$ \\
54753.8324 & 0.4956 & 7.697 & $-$ & $-$ \\
54757.8274 & 0.5919 & 7.943 & $-$ & $-$ \\
54761.7860 & 0.6872 & 8.230 & $-$ & $-$ \\
54765.8078 & 0.7841 & 8.504 & $-$ & $-$ \\
54769.8674 & 0.8819 & 8.714 & $-$ & $-$ \\
54773.8301 & 0.9774 & 8.843 & $-$ & $-$ \\
54781.8260 & 0.1700 & 9.290 & $-$ & $-$ \\
54786.7894 & 0.2896 & 9.420 & $-$ & $-$ \\
54790.7745 & 0.3856 & 8.677 & $-$ & $-$ \\
54794.8251 & 0.4831 & 7.737 & $-$ & $-$ \\
54798.7928 & 0.5787 & 7.889 & $-$ & $-$ \\
54802.7976 & 0.6752 & 8.192 & $-$ & $-$ \\
54806.7729 & 0.7710 & 8.446 & $-$ & $-$ \\
54810.7922 & 0.8678 & 8.634 & $-$ & $-$ \\
54814.8080 & 0.9645 & 8.848 & $-$ & $-$ \\
54818.7459 & 0.0594 & 9.037 & $-$ & $-$ \\
54822.7881 & 0.1568 & 9.281 & $-$ & $-$ \\
54826.8193 & 0.2539 & 9.434 & $-$ & $-$ \\
54834.8558 & 0.4475 & 7.930 & $-$ & $-$ \\
54838.8136 & 0.5428 & 7.805 & $-$ & $-$ \\
54842.7421 & 0.6375 & 8.076 & $-$ & $-$ \\
54850.6641 & 0.8283 & 8.558 & $-$ & $-$ \\
54854.7633 & 0.9270 & 8.777 & $-$ & $-$ \\
54858.8023 & 0.0243 & 8.933 & $-$ & $-$ \\
54862.7753 & 0.1200 & 9.191 & $-$ & $-$ \\
54870.7043 & 0.3111 & 9.336 & $-$ & $-$ \\
54874.6945 & 0.4072 & 8.426 & $-$ & $-$ \\
54878.6485 & 0.5024 & 7.742 & $-$ & $-$ \\
54882.7127 & 0.6003 & 7.937 & $-$ & $-$ \\
54887.6628 & 0.7196 & 8.272 & $-$ & $-$ \\
54891.7283 & 0.8175 & 8.557 & $-$ & $-$ \\
54893.7053 & 0.8651 & 8.636 & $-$ & $-$ \\
54894.6760 & 0.8885 & 8.686 & $-$ & $-$ \\
54898.6346 & 0.9839 & 8.878 & $-$ & $-$ \\
54900.6782 & 0.0331 & 8.981 & $-$ & $-$ \\
54905.6835 & 0.1537 & 9.239 & $-$ & $-$ \\
54909.6473 & 0.2492 & 9.388 & $-$ & $-$ \\
54913.5922 & 0.3442 & 9.119 & $-$ & $-$ \\
54918.6438 & 0.4659 & 7.795 & $-$ & $-$ \\
54922.6326 & 0.5620 & 7.823 & $-$ & $-$ \\
54927.5957 & 0.6816 & 8.212 & $-$ & $-$ \\
54931.6133 & 0.7783 & 8.455 & $-$ & $-$ \\
54934.5792 & 0.8498 & 8.600 & $-$ & $-$ \\
54938.6068 & 0.9468 & 8.796 & $-$ & $-$ \\
54941.6177 & 0.0193 & 8.945 & $-$ & $-$ \\
54947.5678 & 0.1627 & 9.280 & $-$ & $-$ \\
54951.5816 & 0.2594 & 9.358 & $-$ & $-$ \\
54955.5251 & 0.3544 & 9.042 & $-$ & $-$ \\
54959.5129 & 0.4504 & 7.861 & $-$ & $-$ \\
54963.5287 & 0.5472 & 7.814 & $-$ & $-$ \\
54965.5290 & 0.5954 & 7.972 & $-$ & $-$ \\
54967.5193 & 0.6433 & 8.090 & $-$ & $-$ \\
54971.5113 & 0.7395 & 8.331 & $-$ & $-$ \\
54973.4828 & 0.7870 & 8.458 & $-$ & $-$ \\
54978.4828 & 0.9074 & 8.727 & $-$ & $-$ \\
54983.4894 & 0.0280 & 8.963 & $-$ & $-$ \\
54987.4693 & 0.1239 & 9.120 & $-$ & $-$ \\
54991.4909 & 0.2208 & 9.351 & $-$ & $-$ \\
54995.4676 & 0.3166 & 9.296 & $-$ & $-$ \\
55085.8831 & 0.4947 & 7.701 & $-$ & $-$ \\
55089.8848 & 0.5911 & 7.910 & $-$ & $-$ \\
55093.8648 & 0.6870 & 8.218 & $-$ & $-$ \\
55097.8738 & 0.7835 & 8.489 & $-$ & $-$ \\
55103.8741 & 0.9281 & 8.738 & $-$ & $-$ \\
55109.8788 & 0.0727 & 9.079 & $-$ & $-$ \\
55113.8235 & 0.1678 & 9.256 & $-$ & $-$ \\
55120.8393 & 0.3368 & 9.185 & $-$ & $-$ \\
55123.8805 & 0.4100 & 8.325 & $-$ & $-$ \\
55127.7931 & 0.5043 & 7.741 & $-$ & $-$ \\
55134.8193 & 0.6735 & 8.204 & $-$ & $-$ \\
55138.8452 & 0.7705 & 8.443 & $-$ & $-$ \\
55143.7863 & 0.8896 & 8.687 & $-$ & $-$ \\
55147.8228 & 0.9868 & 8.885 & $-$ & $-$ \\
55151.8117 & 0.0829 & 9.079 & $-$ & $-$ \\
55155.8085 & 0.1792 & 9.292 & $-$ & $-$ \\
55159.8390 & 0.2763 & 9.385 & $-$ & $-$ \\
55163.8296 & 0.3724 & 8.785 & $-$ & $-$ \\
55164.8319 & 0.3965 & 8.480 & $-$ & $-$ \\
55177.7902 & 0.7087 & 8.281 & $-$ & $-$ \\
55186.7513 & 0.9246 & 8.753 & $-$ & $-$ \\
55190.7915 & 0.0219 & 8.945 & $-$ & $-$ \\
55194.8218 & 0.1190 & 9.168 & $-$ & $-$ \\
55199.7529 & 0.2378 & 9.412 & $-$ & $-$ \\
55203.7830 & 0.3349 & 9.176 & $-$ & $-$ \\
55208.7977 & 0.4557 & 7.843 & $-$ & $-$ \\
55212.7897 & 0.5518 & 7.806 & $-$ & $-$ \\
55216.7254 & 0.6466 & 8.109 & $-$ & $-$ \\
55220.7906 & 0.7446 & 8.355 & $-$ & $-$ \\
55224.7630 & 0.8403 & 8.580 & $-$ & $-$ \\
55228.7727 & 0.9369 & 8.794 & $-$ & $-$ \\
55232.7692 & 0.0331 & 9.043 & $-$ & $-$ \\
55235.7519 & 0.1050 & 9.163 & $-$ & $-$ \\
55240.7350 & 0.2250 & 9.394 & $-$ & $-$ \\
55244.7295 & 0.3213 & 9.301 & $-$ & $-$ \\
55248.6259 & 0.4151 & 8.282 & $-$ & $-$ \\
55252.6934 & 0.5131 & 7.726 & $-$ & $-$ \\
55258.6595 & 0.6568 & 8.129 & $-$ & $-$ \\
55261.6721 & 0.7294 & 8.342 & $-$ & $-$ \\
55261.7068 & 0.7302 & 8.328 & $-$ & $-$ \\
55263.7155 & 0.7786 & 8.457 & 6.943 & $-$ \\
55263.7161 & 0.7786 & 8.469 & 6.928 & $-$ \\
55267.6559 & 0.8736 & 8.673 & $-$ & $-$ \\
55267.6564 & 0.8736 & 8.661 & $-$ & $-$ \\
55271.6236 & 0.9691 & 8.874 & $-$ & $-$ \\
55271.6242 & 0.9691 & 8.877 & $-$ & $-$ \\
55279.7040 & 0.1638 & 9.243 & 7.511 & 6.642 \\
55279.7045 & 0.1638 & 9.259 & 7.523 & 6.651 \\
55283.5912 & 0.2574 & 9.416 & 7.689 & $-$ \\
55283.5919 & 0.2574 & 9.424 & 7.666 & $-$ \\
55286.6265 & 0.3305 & 9.243 & 7.589 & $-$ \\
55286.6272 & 0.3306 & 9.224 & 7.586 & $-$ \\
55290.6291 & 0.4270 & 8.126 & 6.857 & $-$ \\
55290.6298 & 0.4270 & 8.131 & 6.871 & $-$ \\
55293.5741 & 0.4979 & 7.724 & $-$ & $-$ \\
55293.5747 & 0.4979 & 7.716 & $-$ & $-$ \\
55299.5718 & 0.6424 & 8.082 & $-$ & $-$ \\
55299.5724 & 0.6424 & 8.087 & $-$ & $-$ \\
55303.5526 & 0.7383 & 8.367 & 6.850 & 6.151 \\
55303.5532 & 0.7383 & 8.338 & 6.889 & 6.138 \\
55307.5114 & 0.8337 & 8.581 & $-$ & $-$ \\
55307.5120 & 0.8337 & 8.580 & $-$ & $-$ \\
55312.5013 & 0.9539 & 8.858 & 7.186 & $-$ \\
55312.5020 & 0.9539 & 8.854 & 7.196 & $-$ \\
55317.5405 & 0.0753 & 9.099 & 7.382 & 6.550 \\
55317.5410 & 0.0753 & 9.104 & 7.375 & 6.537 \\
55321.5120 & 0.1709 & 9.313 & 7.561 & 6.708 \\
55321.5126 & 0.1709 & 9.306 & 7.563 & $-$ \\
55325.5007 & 0.2670 & 9.426 & 7.682 & $-$ \\
55325.5014 & 0.2670 & 9.435 & 7.690 & $-$ \\
55335.4743 & 0.5073 & 7.698 & $-$ & $-$ \\
55335.4749 & 0.5073 & 7.706 & $-$ & $-$ \\
55340.5009 & 0.6284 & 8.071 & 6.717 & 6.053 \\
55340.5015 & 0.6284 & 8.095 & 6.753 & 6.081 \\
55343.4582 & 0.6996 & 8.273 & $-$ & $-$ \\
55343.4587 & 0.6996 & 8.263 & $-$ & $-$ \\
55348.4601 & 0.8201 & 8.604 & 6.990 & $-$ \\
55348.4608 & 0.8201 & 8.557 & 6.995 & $-$ \\
55352.4589 & 0.9164 & 8.786 & 7.143 & $-$ \\
55352.4595 & 0.9165 & 8.778 & 7.141 & $-$ \\
55355.4868 & 0.9894 & 8.947 & 7.243 & $-$ \\
55355.4874 & 0.9894 & 8.917 & 7.253 & $-$ \\
55359.4794 & 0.0856 & 9.138 & 7.414 & 6.551 \\
55359.4799 & 0.0856 & 9.081 & 7.400 & 6.535 \\
55365.4575 & 0.2296 & 9.393 & 7.666 & 6.787 \\
55365.4581 & 0.2296 & 9.399 & 7.666 & 6.788 \\
55369.4463 & 0.3257 & 9.289 & 7.607 & 6.796 \\
55369.4469 & 0.3257 & 9.306 & 7.619 & 6.804 \\
55374.4683 & 0.4466 & 7.904 & 6.676 & 6.029 \\
55374.4689 & 0.4467 & 7.889 & 6.642 & 6.104 \\
55378.4333 & 0.5422 & 7.739 & $-$ & $-$ \\
55378.4339 & 0.5422 & 7.786 & $-$ & $-$ \\
55430.9213 & 0.8066 & 8.537 & 6.971 & 6.208 \\
55430.9219 & 0.8066 & 8.565 & 6.965 & 6.185 \\
55431.9315 & 0.8309 & 8.557 & 7.037 & 6.248 \\
55431.9320 & 0.8309 & 8.598 & 7.076 & 6.119 \\
55435.9023 & 0.9266 & 8.786 & 7.138 & $-$ \\
55435.9029 & 0.9266 & 8.809 & 7.156 & $-$ \\
55438.9121 & 0.9991 & 8.974 & 7.254 & $-$ \\
55438.9127 & 0.9991 & 8.938 & 7.243 & $-$ \\
55446.9056 & 0.1916 & 9.333 & 7.604 & 6.732 \\
55446.9062 & 0.1917 & 9.301 & 7.573 & $-$ \\
55453.8738 & 0.3595 & 8.919 & 7.398 & 6.585 \\
55453.8743 & 0.3595 & 8.897 & 7.397 & 6.598 \\
55456.9115 & 0.4327 & 7.983 & 6.787 & $-$ \\
55456.9123 & 0.4327 & 8.004 & 6.782 & $-$ \\
55459.8894 & 0.5044 & 7.726 & 6.556 & 5.941 \\
55459.8900 & 0.5044 & 7.738 & 6.551 & $-$ \\
55464.8508 & 0.6239 & 8.088 & 6.729 & $-$ \\
55464.8515 & 0.6240 & 8.048 & 6.716 & $-$ \\
55468.8700 & 0.7208 & 8.361 & 6.877 & $-$ \\
55468.8707 & 0.7208 & 8.359 & 6.843 & $-$ \\
55471.8227 & 0.7919 & 8.523 & 6.971 & $-$ \\
55471.8234 & 0.7919 & 8.517 & 6.938 & $-$ \\
55476.8630 & 0.9133 & 8.778 & $-$ & $-$ \\
55476.8636 & 0.9133 & 8.766 & $-$ & $-$ \\
55480.8274 & 0.0088 & 8.954 & 7.286 & 6.441 \\
55480.8280 & 0.0088 & 8.977 & 7.297 & $-$ \\
55484.8926 & 0.1067 & 9.165 & 7.469 & 6.598 \\
55484.8931 & 0.1068 & 9.174 & 7.469 & 6.593 \\
55487.8774 & 0.1786 & 9.329 & 7.583 & $-$ \\
55487.8780 & 0.1787 & 9.317 & 7.582 & $-$ \\
55490.8463 & 0.2502 & 9.408 & 7.685 & $-$ \\
55490.8469 & 0.2502 & 9.405 & 7.672 & $-$ \\
55494.8314 & 0.3462 & 9.091 & 7.506 & 6.703 \\
55494.8320 & 0.3462 & 9.079 & 7.505 & 6.655 \\
55499.8628 & 0.4674 & 7.758 & 6.588 & $-$ \\
55499.8634 & 0.4674 & 7.744 & $-$ & $-$ \\
55503.7804 & 0.5617 & 7.845 & $-$ & $-$ \\
55503.7809 & 0.5618 & 7.848 & $-$ & $-$ \\
55505.8172 & 0.6108 & 8.008 & $-$ & $-$ \\
55505.8178 & 0.6108 & 7.985 & $-$ & $-$ \\
55510.8182 & 0.7313 & 8.349 & $-$ & $-$ \\
55510.8188 & 0.7313 & 8.330 & $-$ & $-$ \\
55514.8232 & 0.8278 & 8.583 & $-$ & $-$ \\
55514.8238 & 0.8278 & 8.569 & $-$ & $-$ \\
55519.8138 & 0.9480 & 8.807 & 7.181 & $-$ \\
55519.8145 & 0.9480 & 8.846 & 7.192 & $-$ \\
55522.7774 & 0.0194 & 8.980 & 7.287 & $-$ \\
55522.7780 & 0.0194 & 8.982 & 7.281 & $-$ \\
55527.8711 & 0.1421 & 9.260 & 7.525 & $-$ \\
55527.8718 & 0.1421 & 9.254 & 7.530 & $-$ \\
55531.8225 & 0.2373 & 9.400 & 7.684 & 6.819 \\
55531.8231 & 0.2373 & 9.445 & 7.677 & $-$ \\
55535.8033 & 0.3332 & 9.230 & 7.597 & 6.763 \\
55535.8039 & 0.3332 & 9.233 & 7.604 & $-$ \\
55540.8641 & 0.4551 & 7.774 & $-$ & $-$ \\
55540.8646 & 0.4551 & 7.801 & $-$ & $-$ \\
55541.7981 & 0.4776 & 7.698 & $-$ & $-$ \\
55541.7986 & 0.4776 & 7.700 & $-$ & $-$ \\
55542.8176 & 0.5021 & 7.687 & $-$ & $-$ \\
55542.8182 & 0.5022 & 7.663 & $-$ & $-$ \\
55543.7799 & 0.5253 & 7.718 & $-$ & $-$ \\
55543.7805 & 0.5253 & 7.731 & $-$ & $-$ \\
55545.8222 & 0.5745 & 7.867 & $-$ & $-$ \\
55545.8227 & 0.5745 & 7.873 & $-$ & $-$ \\
55546.8471 & 0.5992 & 7.930 & 6.624 & $-$ \\
55546.8477 & 0.5992 & 7.945 & 6.653 & $-$ \\
55547.8008 & 0.6222 & 8.034 & 6.690 & $-$ \\
55547.8014 & 0.6222 & 8.017 & 6.683 & $-$ \\
55548.7864 & 0.6459 & 8.085 & $-$ & $-$ \\
55548.7869 & 0.6459 & 8.091 & $-$ & $-$ \\
55549.8240 & 0.6709 & 8.165 & 6.821 & $-$ \\
55549.8247 & 0.6709 & 8.142 & $-$ & $-$ \\
55550.7313 & 0.6928 & 8.233 & $-$ & $-$ \\
55550.7319 & 0.6928 & 8.191 & $-$ & $-$ \\
55551.7763 & 0.7180 & 8.278 & 6.838 & $-$ \\
55551.7769 & 0.7180 & 8.300 & $-$ & $-$ \\
55577.7883 & 0.3446 & 9.122 & 7.526 & $-$ \\
55577.7890 & 0.3446 & 9.137 & 7.520 & $-$ \\
55578.7657 & 0.3681 & 8.861 & 7.368 & $-$ \\
55578.7664 & 0.3681 & 8.865 & 7.362 & $-$ \\
55579.7229 & 0.3912 & 8.586 & 7.180 & $-$ \\
55579.7236 & 0.3912 & 8.564 & $-$ & $-$ \\
55581.7737 & 0.4406 & 7.940 & $-$ & $-$ \\
55581.7743 & 0.4406 & 7.925 & $-$ & $-$ \\
55582.7597 & 0.4643 & 7.762 & $-$ & $-$ \\
55582.7602 & 0.4644 & 7.757 & $-$ & $-$ \\
55583.7457 & 0.4881 & 7.705 & $-$ & $-$ \\
55583.7462 & 0.4881 & 7.680 & $-$ & $-$ \\
55584.7586 & 0.5125 & 7.703 & 6.563 & $-$ \\
55584.7593 & 0.5125 & 7.706 & 6.541 & $-$ \\
55585.7983 & 0.5375 & 7.776 & 6.573 & $-$ \\
55585.7990 & 0.5376 & 7.797 & 6.579 & $-$ \\
55586.7503 & 0.5605 & 7.818 & $-$ & $-$ \\
55586.7509 & 0.5605 & 7.818 & $-$ & $-$ \\
55587.7925 & 0.5856 & 7.904 & 6.642 & 5.993 \\
55587.7930 & 0.5856 & 7.906 & 6.648 & 6.000 \\
\end{supertabular}


\begin{figure}[]
\centering
\includegraphics[width=\hsize]{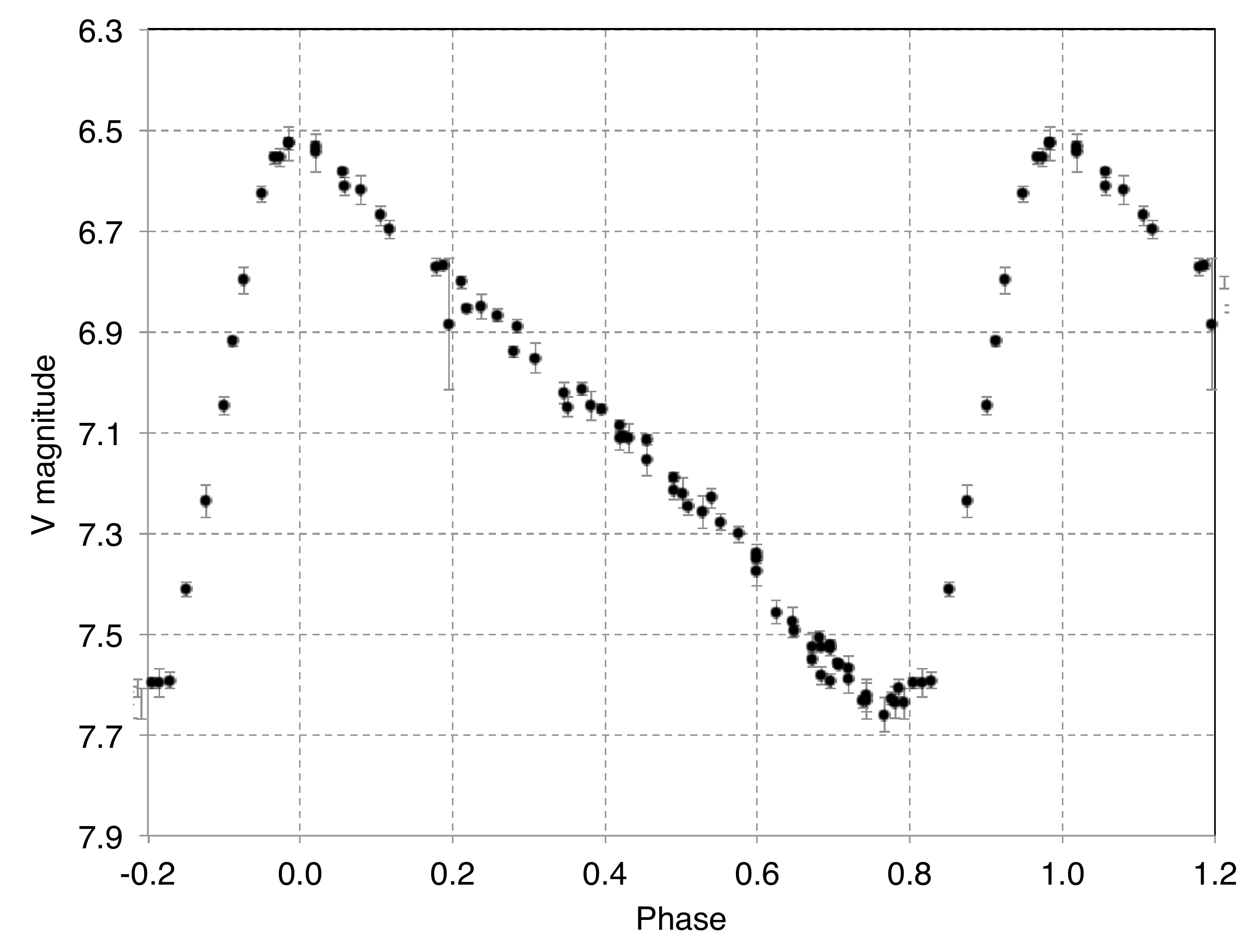}
\caption{AAVSO light curve of RS\,Pup in the Johnson $V$ band.\label{AAVSOphotometry}}
\end{figure}

\twocolumn
\begin{table}
\caption{Photometry of RS\,Pup in the Johnson $V$ band collected by AAVSO observer Neil Butterworth (BIW) from Mt. Louisa, Australia using a transformed DSLR.
\label{tab-aavso}}
\end{table}
\tablehead{\hline \hline \noalign{\smallskip} $\mathrm{JD}_\odot -2.4 \times 10^6$ & Phase & $V$ \\ \hline \noalign{\smallskip}}
\tabletail{\hline}
\begin{supertabular}{cccc}
56404.8860 & $ 0.7380 $ & $ 7.634 \pm 0.013 $ \\
56406.8808 & $ 0.7861 $ & $ 7.608 \pm 0.017 $ \\
56423.8730 & $ 0.1956 $ & $ 6.886 \pm 0.131 $ \\
56424.8695 & $ 0.2196 $ & $ 6.854 \pm 0.008 $ \\
56436.8738 & $ 0.5089 $ & $ 7.248 \pm 0.016 $ \\
56455.8664 & $ 0.9666 $ & $ 6.556 \pm 0.012 $ \\
56468.8603 & $ 0.2797 $ & $ 6.941 \pm 0.011 $ \\
56471.8580 & $ 0.3519 $ & $ 7.050 \pm 0.020 $ \\
56474.8600 & $ 0.4243 $ & $ 7.108 \pm 0.012 $ \\
56702.9424 & $ 0.9256 $ & $ 6.799 \pm 0.026 $ \\
56710.9415 & $ 0.1183 $ & $ 6.698 \pm 0.018 $ \\
56731.9182 & $ 0.6238 $ & $ 7.457 \pm 0.024 $ \\
56732.9219 & $ 0.6480 $ & $ 7.493 \pm 0.011 $ \\
56733.9270 & $ 0.6722 $ & $ 7.552 \pm 0.014 $ \\
56734.9298 & $ 0.6964 $ & $ 7.593 \pm 0.015 $ \\
56749.9202 & $ 0.0576 $ & $ 6.612 \pm 0.019 $ \\
56751.9200 & $ 0.1058 $ & $ 6.670 \pm 0.020 $ \\
56754.9375 & $ 0.1785 $ & $ 6.772 \pm 0.017 $ \\
56761.9075 & $ 0.3465 $ & $ 7.022 \pm 0.021 $ \\
56764.8938 & $ 0.4184 $ & $ 7.110 \pm 0.025 $ \\
56767.9004 & $ 0.4909 $ & $ 7.215 \pm 0.019 $ \\
56775.8966 & $ 0.6836 $ & $ 7.583 \pm 0.017 $ \\
56781.9189 & $ 0.8287 $ & $ 7.592 \pm 0.016 $ \\
56784.8895 & $ 0.9003 $ & $ 7.047 \pm 0.019 $ \\
56787.9128 & $ 0.9731 $ & $ 6.555 \pm 0.019 $ \\
56798.9200 & $ 0.2384 $ & $ 6.850 \pm 0.025 $ \\
56801.8710 & $ 0.3095 $ & $ 6.953 \pm 0.029 $ \\
56819.8598 & $ 0.7430 $ & $ 7.633 \pm 0.037 $ \\
56820.8664 & $ 0.7673 $ & $ 7.660 \pm 0.034 $ \\
57095.9079 & $ 0.3948 $ & $ 7.053 \pm 0.009 $ \\
57096.9064 & $ 0.4188 $ & $ 7.087 \pm 0.011 $ \\
57099.9117 & $ 0.4912 $ & $ 7.189 \pm 0.008 $ \\
57107.9034 & $ 0.6838 $ & $ 7.527 \pm 0.006 $ \\
57108.9093 & $ 0.7081 $ & $ 7.563 \pm 0.006 $ \\
57111.9094 & $ 0.7804 $ & $ 7.635 \pm 0.032 $ \\
57136.8846 & $ 0.3822 $ & $ 7.048 \pm 0.028 $ \\
57138.8867 & $ 0.4304 $ & $ 7.111 \pm 0.029 $ \\
57139.8798 & $ 0.4543 $ & $ 7.153 \pm 0.033 $ \\
57141.8801 & $ 0.5026 $ & $ 7.221 \pm 0.029 $ \\
57142.8984 & $ 0.5271 $ & $ 7.258 \pm 0.032 $ \\
57145.8928 & $ 0.5993 $ & $ 7.377 \pm 0.029 $ \\
57147.8809 & $ 0.6472 $ & $ 7.477 \pm 0.030 $ \\
57148.8779 & $ 0.6712 $ & $ 7.527 \pm 0.031 $ \\
57150.8790 & $ 0.7194 $ & $ 7.589 \pm 0.028 $ \\
57151.8755 & $ 0.7434 $ & $ 7.623 \pm 0.032 $ \\
57153.8838 & $ 0.7918 $ & $ 7.638 \pm 0.031 $ \\
57154.8876 & $ 0.8160 $ & $ 7.596 \pm 0.029 $ \\
57161.8776 & $ 0.9844 $ & $ 6.528 \pm 0.033 $ \\
57165.8708 & $ 0.0806 $ & $ 6.619 \pm 0.030 $ \\
57198.8614 & $ 0.8757 $ & $ 7.236 \pm 0.033 $ \\
57204.8609 & $ 0.0201 $ & $ 6.546 \pm 0.038 $ \\
57392.9682 & $ 0.5517 $ & $ 7.278 \pm 0.016 $ \\
57393.9915 & $ 0.5764 $ & $ 7.302 \pm 0.017 $ \\
57394.9576 & $ 0.5996 $ & $ 7.339 \pm 0.017 $ \\
57398.9627 & $ 0.6962 $ & $ 7.529 \pm 0.014 $ \\
57399.9590 & $ 0.7202 $ & $ 7.567 \pm 0.025 $ \\
57402.2453 & $ 0.7753 $ & $ 7.628 \pm 0.012 $ \\
57410.9215 & $ 0.9842 $ & $ 6.528 \pm 0.012 $ \\
57433.9313 & $ 0.5388 $ & $ 7.231 \pm 0.019 $ \\
57444.9741 & $ 0.8049 $ & $ 7.597 \pm 0.012 $ \\
57446.8990 & $ 0.8513 $ & $ 7.411 \pm 0.015 $ \\
57471.9630 & $ 0.4552 $ & $ 7.115 \pm 0.010 $ \\
57477.9219 & $ 0.5988 $ & $ 7.351 \pm 0.009 $ \\
57481.9689 & $ 0.6963 $ & $ 7.522 \pm 0.009 $ \\
57490.9058 & $ 0.9117 $ & $ 6.920 \pm 0.009 $ \\
57496.9034 & $ 0.0561 $ & $ 6.582 \pm 0.004 $ \\
57509.9115 & $ 0.3696 $ & $ 7.014 \pm 0.012 $ \\
57522.8794 & $ 0.6821 $ & $ 7.508 \pm 0.015 $ \\
57523.8833 & $ 0.7063 $ & $ 7.557 \pm 0.008 $ \\
57533.9424 & $ 0.9487 $ & $ 6.627 \pm 0.016 $ \\
57536.8819 & $ 0.0194 $ & $ 6.532 \pm 0.009 $ \\
57543.8699 & $ 0.1878 $ & $ 6.768 \pm 0.010 $ \\
57544.8855 & $ 0.2123 $ & $ 6.803 \pm 0.012 $ \\
57546.8602 & $ 0.2598 $ & $ 6.868 \pm 0.013 $ \\
57547.8634 & $ 0.2840 $ & $ 6.889 \pm 0.014 $ \\
\end{supertabular}